\documentclass[letter]{JHEP3}

\JHEPspecialurl{http://jhep.sissa.it/JOURNAL/JHEP3.tar.gz}

\usepackage{graphicx}
\usepackage{amsmath}
\usepackage{epsfig,multicol,bbm}

\newcommand\fverb{\setbox\fverbbox=\hbox\bgroup\verb}
\newcommand\fverbdo{\egroup\medskip\noindent%
			\fbox{\unhbox\fverbbox}\ }
\newcommand\fverbit{\egroup\item[\fbox{\unhbox\fverbbox}]}
\newbox\fverbbox

\title{Inelastic Dark Matter, Non-Standard Halos and the DAMA/LIBRA Results}

\author{John March-Russell\\
	Rudolf Peierls Centre for Theoretical Physics, University of Oxford, 1 Keble Road, Oxford, OX1 3NP, UK\\
	E-mail: \email{jmr@thphys.ox.ac.uk}}
\author{Christopher McCabe\\
	Rudolf Peierls Centre for Theoretical Physics, University of Oxford, 1 Keble Road, Oxford, OX1 3NP, UK\\
	E-mail: \email{mccabe@thphys.ox.ac.uk}}
\author{Matthew McCullough\\
	Rudolf Peierls Centre for Theoretical Physics, University of Oxford, 1 Keble Road, Oxford, OX1 3NP, UK\\
	E-mail: \email{mccull@thphys.ox.ac.uk}}

\preprint{OUTP-08 19 P}	% OR: \preprint{Aaaa/Mm/Yy\\Aaa-aa/Nnnnnn}
\abstract{The DAMA collaboration have claimed to detect particle dark matter (DM) via an annual modulation in their observed recoil event rate.  This appears to be in strong disagreement with the null results of other experiments if interpreted in terms of elastic DM scattering, while agreement for a small region of parameter space is possible for inelastic DM (iDM) due to the altered kinematics of the collision.  To date most analyses assume a simple galactic halo DM velocity distribution, the Standard Halo Model, but direct experimental support for the SHM is severely lacking and theoretical studies indicate possible significant differences.   We investigate the dependence of DAMA and the other direct detection experiments on the local DM velocity distribution, utilizing the results of the Via Lactea and Dark Disc numerical simulations.  We also investigate effects of varying the solar circular velocity, the DM escape velocity, and the DAMA quenching factor within experimental limits.
Our data set includes the latest ZEPLIN-III results, as well as full publicly available data sets.   Due to the more sensitive dependence of the inelastic cross section on the velocity distribution, we find that with Via Lactea the DAMA results can be consistent with all other experiments over an enlarged region of iDM parameter space, with higher mass particles being preferred, while Dark Disc does not lead to an improvement.  A definitive test of DAMA for iDM requires heavy element detectors.
}


\begin{document}

\section{Introduction}
First-hand information on the nature of the dark matter (DM) in our universe may be revealed by one or more of the current generation of direct detection experiments.  The DAMA collaboration have measured an annual modulation in their scintillation event rate, initially with the DAMA/NaI set-up \cite{Bernabei:2005hj}, and more recently with the DAMA/LIBRA 
configuration \cite{Bernabei:2008yi}, which they interpret as being due to the modulation in the DM-nuclear scattering event rate following from the annual variation of the Earth's velocity through the DM in our galaxy.  At face value, this measurement appears to be in disagreement with other direct detection experiments if we make the usual assumptions about the nature of the DM; namely that the DM is a weakly interacting massive particle (WIMP) that elastically scatters off the target material with a spin independent interaction.

Inelastic dark matter (iDM) \cite{TuckerSmith:2001hy} was proposed as a way to reconcile the positive result from DAMA/NaI with the null result from germanium based detectors.  In the iDM scenario, WIMP-nucleon elastic scattering is suppressed, while inelastic scattering from a ground-state WIMP to a slightly higher mass excited WIMP is allowed and dominates the recoil event rate.  As we summarize in Section \ref{basics}, the kinematics of the recoil scattering are changed by the inelastic nature of the collision, and this can bring the DAMA results closer to agreement with the other experiments for three principal reasons:
\begin{itemize}
\item Heavier nuclei are favoured.
\item The recoil spectrum is changed at low energies.
\item The ratio of modulated to unmodulated signal is higher.
\end{itemize}
Nevertheless the other direct detection experiments still strongly constrain the DM interpretation of DAMA/LIBRA, with, apparently, only a relatively small region of parameter space being allowed \cite{Chang:2008gd}.    

One unchecked assumption that goes into the analysis is that the velocity distribution of the DM in the galactic rest frame is well described by so-called Standard Halo Model (SHM).   The SHM assumes that in the galactic frame the WIMP distribution is an isotropic isothermal sphere, which leads to an essentially structureless isotropic Maxwell-Boltzmann velocity distribution with dispersion set by the local circular velocity.
This is a questionable assumption: Little is known about the phase-space distribution of DM on the scales relevant for direct detection experiments, and as we outline in Section \ref{haloes} numerical simulations
of DM distributions for Milky-Way-like galaxies lead to results differing from the SHM in potentially significant ways, especially for iDM which is more sensitive to the DM velocity distribution.

In this paper, we compare the results obtained using the SHM to those obtained using two recent computer simulations of the DM distribution in a Milky-Way-like galaxy:
\begin{itemize}
\item Via Lactea \cite{Diemand:2006ik} - a Milky Way size DM halo distribution containing 234 million particles.
\item Dark Disc \cite{Read:2008fh} - a simulation which contains in addition to the SHM, a slowly rotating disc of DM
\end{itemize}
With these less idealised halo models, we find that the iDM scenario allows the DAMA region to be consistent with all other experiments, and that in the Via Lactea distribution, more parameter space can be opened up at high WIMP masses, compared to the SHM.   After reviewing the details of each experiment and describing how we calculate the allowed DAMA region and exclusion curves for the null experiments in Section \ref{exper}, we
present our results together with their physical interpretation in Section \ref{Results}.

\section{Inelastic dark matter}\label{basics}

\subsection{Review of inelastic dark matter}

If the dark matter particle can only scatter off nuclei by making a transition to a heavier state then the altered kinematics of the interaction can lead to significant changes in detection rates for different detectors \cite{TuckerSmith:2001hy, Chang:2008gd, TuckerSmith:2004jv}.  If we call the two dark matter states $\chi_-$ and $\chi_+$, with mass splitting $\delta \equiv M_{\chi_{+}} - M_{\chi_{-}} \sim \text{O(100 keV)}$, then the minimum velocity to scatter off a nucleus and impart an energy $E_R$ is given by;
\begin{equation}
v_{min} = c\sqrt{{1} \over {2 M_N E_R}} \left({{M_N E_R} \over {\mu}} + \delta \right) ,
\label{vmin}
\end{equation}
where $\mu$ is the reduced mass of the WIMP-nucleus system, and $M_N$ is the nucleus mass.
There are a variety of particle physics models that lead to such phenomenology \cite{TuckerSmith:2001hy,ArkaniHamed:2000bq,MarchRussell:2004uf}.  The models will not be our concern
in this paper, rather our focus will be upon the consequences of iDM for direct detection experiments.

\begin{figure}[h]
\centering
\includegraphics[height=1.8in]{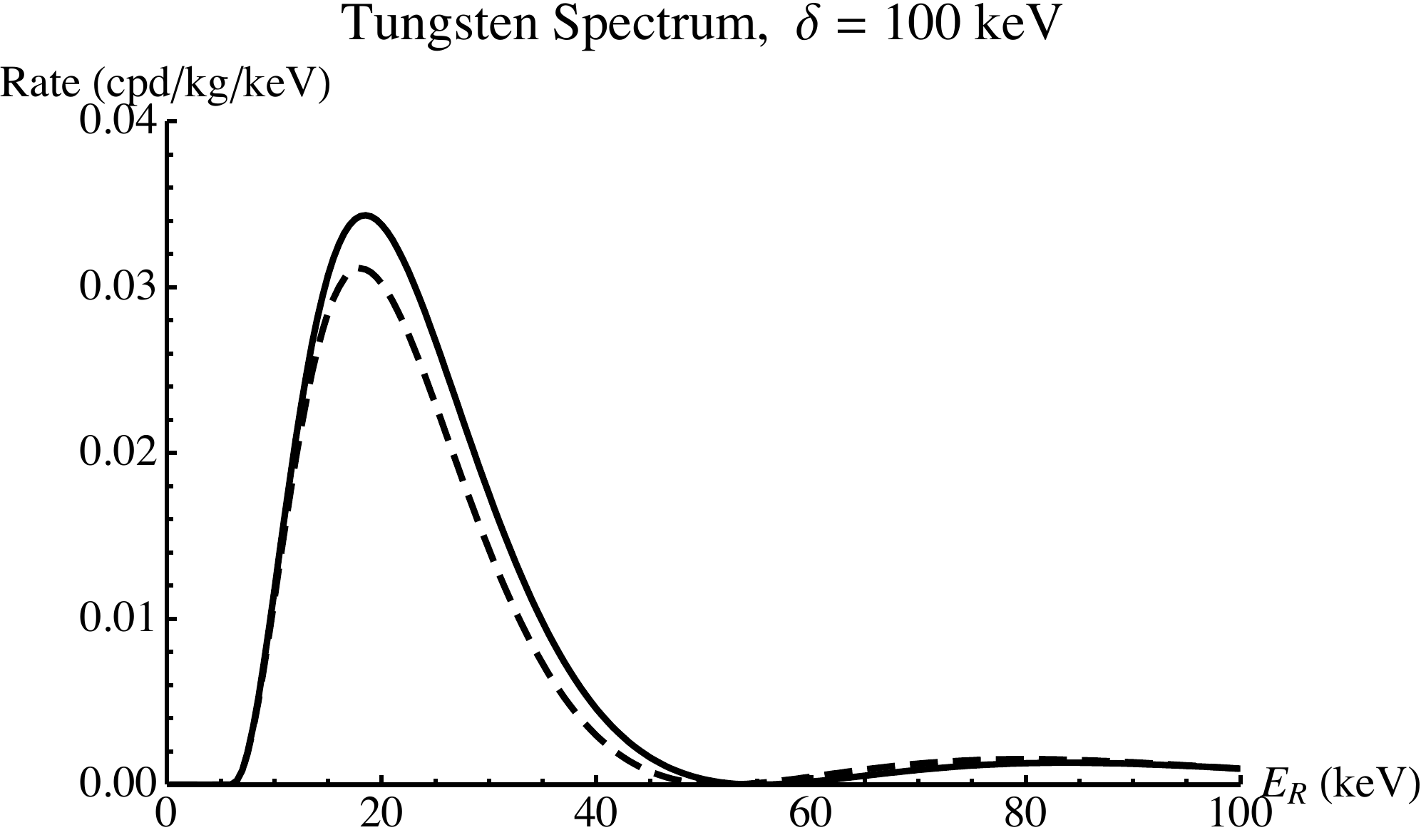}\includegraphics[height=1.8in]{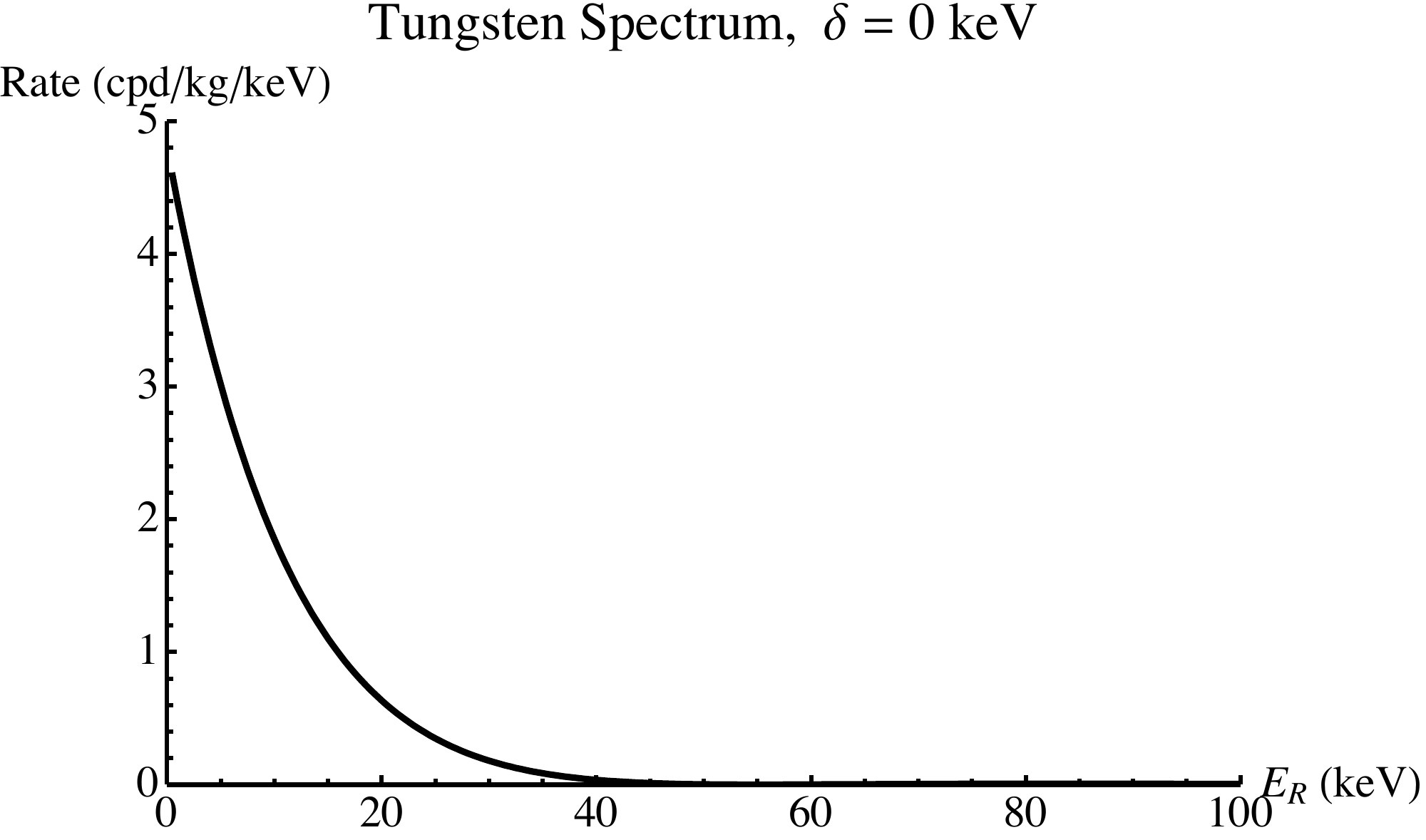}\caption{Recoil energy spectrum for scattering off of tungsten.   On the left the inelastic scattering rates are shown for the two choices of form factor: the Fermi Two-Parameter (dashed) and Helm (solid) form factor.  Both choices show that iDM leads to a suppression of low-energy events.    The right panel shows the recoil energy spectrum for elastic scattering; in this case the difference between the two form factors is negligible.  For elastic scattering the recoil spectrum peaks at low $E_R$.  All calculations assume $M_\chi = 200 \: $GeV, $\sigma_n = 10^{-40} $ cm$^{2}$ and the SHM velocity distribution  truncated at $v_{esc} = 500$ km/s.}\label{fig:formplot}
\end{figure}

The differential event rate for WIMP nucleus scattering on a given element, as a function of recoil energy $E_R$, is given by, \cite{Lewin:1995rx}:
\begin{equation}
{dR \over dE_R} = N_T M_N {\rho_{\chi} {\sigma}_n \over 2 m_{\chi} {{\mu}^2_{ne}}} {(f_p Z + f_n (A-Z))^2 \over {f_n}^2} F^2[E_R] \int^\infty_{v_{min}[E_R]} {f(v) \over v} dv ,
\label{eventrate}
\end{equation}
where, as in \cite{Chang:2008gd}, $N_T$ is the number of target nuclei, $\rho_\chi$ is the local dark matter density, taken as $0.3 \text{ GeV}/\text{cm}^3$,  ${\mu}_{ne}$ the reduced mass of the WIMP and nucleon, and ${\sigma}_n$ is the WIMP-neutron cross section. The factors $f_n$ and $f_p$
parameterize the relative scattering strength off neutrons and protons.
For simplicity and to maintain model-independence we assume that scattering off neutrons and protons is the same and take $f_n = f_p = 1$. (This assumption is not always correct for candidate dark matter particles from specific models.)  $F[E_R]$ is the nuclear form factor, which can be modeled analytically with the Helm form factor \cite{Helm:1956zz} or calculated numerically using the Fermi Two-Parameter form factor \cite{Duda:2006uk}.  In accordance with \cite{Chang:2008gd} we will use the Helm form factor for CDMS II, KIMS, XENON10, ZEPLIN-II, and ZEPLIN-III, and the Fermi Two-Parameter form factor for CRESST-II and DAMA/LIBRA. 
In the first panel of Figure \ref{fig:formplot} we show for tungsten the difference that results between the form factors.

Finally, $f(v)$, is the local velocity distribution of the WIMPs in the Earth rest frame. The default assumption
for the dark matter phase space density is the so-called ``Standard Halo Model"  (SHM) which assumes a local isothermal and isotropic distribution of dark matter leading to a Maxwell-Boltzmann velocity distribution.
Changes in this velocity distribution make major differences to the final differential event rate as calculated from Eq.(\ref{eventrate}), especially for iDM. We emphasize that $f(v)$ is poorly constrained by data and not well understood.  The primary subject of this paper is to investigate the allowed parameter space for iDM taking account of reasonable variations in $f(v)$ motivated by numerical simulations of the DM distribution in our galaxy.

The most important difference between inelastic and standard elastic dark matter is the $\delta$-dependent
increase in the minimum relative velocity for scattering with recoil energy $E_R$, see Eq.(\ref{vmin}).  The important consequences for direct detection experiments are:
\begin{itemize}
\item As only higher velocity portions of the dark matter distribution lead to recoils the overall event rate for a given cross section is lower.
\item The spectrum of events is changed, with the greatest qualitative departure from elastic scattering occurring at low energies.  As one can see from the first panel of Figure \ref{fig:formplot} there is a low energy cut off in the recoil spectrum.  This is not present for elastic scattering as shown in the second panel of Figure \ref{fig:formplot}.
\item Due to the higher minimum velocity the annual modulation as a fraction of the total signal can be much greater.  It is possible that detection of WIMPs occurs solely during the summer months, when the Earth's velocity into the WIMP wind is greatest, leading to a modulation fraction of 100\%. This has important
consequences both for DAMA/LIBRA and XENON10. 
\item As the minimum velocity cut-off, Eq.(\ref{vmin}), is lower for heavier target nuclei, inelastic scattering leads to higher expected detection rates for heavier elements.  This has important consequences for the sensitivity of the CDMS II ($A_{Ge}=72.64$) and CRESST-II ($A_W=183.84$) experiments.
\end{itemize}

\subsection{Overview of consequences of iDM for the various experiments}

The minimum velocity for scattering will be much greater for the CDMS II detector compared to DAMA/LIBRA, where the dominant signal is from scattering off of germanium and iodine respectively, because germanium is much lighter than iodine.  This leads to a weakening of the constraints from CDMS II on the region of parameter space preferred by the DAMA/LIBRA annual modulation signal.  In fact for high enough $\delta$ the expected rate for CDMS II can be consistent with zero, while still allowing signal at DAMA/LIBRA.
The converse effect applies to the CRESST-II experiment, where scattering occurs off of tungsten.  Here we would expect a higher rate for scattering, which leads to CRESST-II setting the most stringent constraints on the DAMA/LIBRA results.  

The three xenon based experiments, XENON10, ZEPLIN-II, and its successor ZEPLIN-III, provide an excellent model-independent test of the DAMA/LIBRA results due to the similarity of xenon and iodine masses ($A_{Xe}=131.293, A_{I}=126.904$). However the events observed at these detectors lead to consistency with the DAMA/LIBRA preferred region.  Also, due to the enhancement of the modulation effect, detectors which took results over the winter period, when signals would be lowest, inevitably set lower constraints than possible if running during the summer months.  This applies in particular to the XENON10 experiment.  Moreover, the low upper-energy limit (30.2 keV) in the analyzed recoil energy spectrum of ZEPLIN-III reduces the sensitivity of this experiment to the iDM scenario, as for a typical $\delta$ of 100 keV the xenon recoil spectrum peaks at $\sim$40 keV.

We return to the details of the experiments in section \ref{exper}.

\section{Dark matter halos}  \label{haloes}

Due to the long range of the gravitational force, and the distribution of matter in our galaxy, one would expect the correct velocity distribution of dark matter particles to deviate from exact Maxwellian and to show some anisotropy.  N-body simulations of large numbers of dark matter particles have shown that the SHM may well be incorrect \cite{Diemand:2006ik,Read:2008fh}. 

Here we will consider three models for the local velocity distribution of particles in the dark matter halo.  We will always take the Earth velocity with respect to the galactic rest frame to be given by $\boldsymbol{v}_{Earth} = \boldsymbol{v}_{\odot} + \boldsymbol{v}_{\oplus}$, where $\boldsymbol{v}_{\odot}$ is the Sun's velocity relative to the galactic rest frame and $\boldsymbol{v}_{\oplus}$ is the Earth's velocity relative to the Sun.  $\boldsymbol{v}_{\odot}$ is the sum of the Sun's peculiar \cite{Dehnen:1997cq} and circular \cite{BinneyBook} velocities:
\begin{eqnarray}
\boldsymbol{v}_{\odot} = 
\left(
\begin {array}{cc}
10.00\\
5.23\\
7.17
\end {array}
\right) \text{km/s} + 
\left(
\begin {array}{cc}
0\\
220\\
0
\end {array}
\right) \text{km/s}
\end{eqnarray}

$\boldsymbol{v}_{\oplus}$ is given by \cite{Lewin:1995rx};

\begin{eqnarray}
\boldsymbol{v}_{\oplus} = 
\langle u_E \rangle (1-e \sin(\lambda(n) - \lambda_0))
\left(
\begin {array}{cc}
\cos(\beta_x) \sin(\lambda(n) - \lambda_x)\\
\cos(\beta_y) \sin(\lambda(n) - \lambda_y)\\
\cos(\beta_z) \sin(\lambda(n) - \lambda_z)
\end {array}
\right) \text{km/s}
\end{eqnarray}
where the Earth's orbit has a mean velocity $\langle u_E \rangle=29.79$ km/s and ellipticity $e = 0.016722$.  The quantities $\beta_i$, $\lambda_i$, define the orientation of the Earth's orbit in galactic coordinates, and $\lambda(n)$ gives the angular position of the Earth's orbit for a given day number $n$, with $n=1$ corresponding to 1st January 2000.  These quantities are given in \cite{Lewin:1995rx}.

The WIMP velocity distribution in the Earth's rest frame is found by performing a Galilean boost from the galactic rest frame distribution:
\begin{equation}
f_{\oplus}(\boldsymbol{v},t) = f_{gal}(\boldsymbol{v} + \boldsymbol{v}_{Earth}(t))
\end{equation}
One feature common to all of the velocity distributions considered here is that they are truncated at the local escape velocity, $f_{gal}(|\boldsymbol{v}| > v_{esc}) = 0$.  There is relatively large error in the known value of the local escape velocity, $498 < v_{esc} < 608 $ km/s (90\%), with a median of $544$ km/s \cite{Smith:2006ym}.  We have taken $v_{esc} = 550$ km/s  in the analysis presented here. In Section \ref{escapevel} we check that varying $v_{esc}$ between the 90\% confidence limits does not change the qualitative features of our results, and therefore the conclusions.

\subsection{The Standard Halo Model}
In the Standard Halo Model (SHM) one assumes a Maxwellian velocity distribution for the dark matter particles, with a local velocity dispersion $\sigma = \overline{v}/\sqrt{2}$, where $\overline{v} = \sqrt{-r \frac{dU(\boldsymbol{r})}{dr}}$ is the local circular velocity, and $U(\boldsymbol{r})$ is the gravitational potential.  This distribution is truncated at the escape velocity according to:
\begin{equation}
f(\boldsymbol{v}) = \left\{ \begin{array}{cr}
  \frac{1}{N} [\exp({-v^2/\overline{v}^2})-\exp({-{v_{esc}^2}/\overline{v}^2})] & v < v_{esc} \\
  0 & v > v_{esc}
       \end{array} \right. 
\end{equation}

\subsection{Via Lactea}

Here we use results for the phase space distribution of dark matter in a Milky-Way-like galaxy derived from a simulation containing 234 million
particles of dark matter and no baryons; Via Lactea \cite{Diemand:2006ik} published in 2006.

In \cite{Fairbairn:2008gz} the velocity distribution of dark matter particles was fitted to the distribution of individual particles from the Via Lactea simulation \cite{Diemand:2006ik}.  The radial and tangential velocity distributions were fitted according to;
\begin{eqnarray}
f(v_R) =  \frac{1}{N_R} \exp \left[{-\biggl(\frac{v_R^2}{\overline{v}^2_R}}\biggr)^{\alpha_R}\right]\\
f(v_T) =  \frac{2 \pi v_T}{N_T} \exp \left[{-\biggl(\frac{v_T^2}{\overline{v}^2_T}}\biggr)^{\alpha_T}\right]
\end{eqnarray}
and the distribution was truncated in the same way as for the SHM.  From Figure 3 in \cite{Fairbairn:2008gz} we extracted the values $\alpha_R \approx 1.09$, and $\overline{v}_R/(\sqrt{-U(\boldsymbol{r}_{0})}) \approx 0.72$ and $\alpha_T \approx 0.73$,  and $\overline{v}_T/(\sqrt{-U(\boldsymbol{r}_{0})}) \approx 0.47$ at our radius from the centre of the galaxy, $r_0\approx  8.5$ kpc. (Although there is uncertainty in our radius from the galactic centre, $r_0 = 8.0 \pm 0.5$ kpc \cite{BinneyBook}, from Figure 3 in \cite{Fairbairn:2008gz} one can see that the values of $\alpha_T$ and $\overline{v}_T$ do not change significantly over the range $7.5 < r_0 < 8.5$ kpc, and the tangential velocity distribution is the main determinant of the event rate).

For completeness we will present results using two values for $\sqrt{-U(\boldsymbol{r}_{0})}$.  In the Via Lactea halo the average value of $\sqrt{-U(\boldsymbol{r})}$ between 7 and 9 kpc is $270\text{km/s}$ \cite{Kuhlen}; we will refer to results using this value as $\text{VL}_{270}$.  Following \cite{Fairbairn:2008gz}, and in order to allow direct comparison between inelastic and elastic scenarios, we also present results using $\sqrt{-U(\boldsymbol{r_0})} = 220\text{km/s}$ \cite{MFairbairn}, and refer to these as $\text{VL}_{220}$.

We believe both values for $\sqrt{-U(\boldsymbol{r}_{0})}$ are worth studying, given that the Milky Way is baryon dominated at the solar radius, so uncertainties will arise from the lack of baryons in any simulation which contains only dark matter particles. It is also important to remember that the velocity distribution extracted from any DM simulation is for a Milky-Way-like galaxy, and not the Milky-Way itself. Therefore it is important to study the effects of reasonable deviations from results predicted by a simulation.

The deviation from the Gaussian distribution and difference in radial and tangential velocity dispersions have been shown in \cite{Fairbairn:2008gz} to affect the expected DAMA/LIBRA modulation signal for elastic dark matter, although not by enough to allow an elastic DM interpretation of the DAMA results.

Due to the smaller tangential velocity dispersion in this model, compared to the SHM, one would expect the results for inelastic scattering to be changed for light nuclei.  This is because the high minimum velocity for scattering on light nuclei leaves only the high velocity part of the distribution detectable.  This smaller velocity dispersion reduces the population of this high velocity region further, leading to a reduced event rate.  This effect is particularly interesting as it arises through the combination of iDM and the Via Lactea halo.   We further discuss this in Section~5.1.

\subsection{Dark Disc}

Previous simulations of the DM phase space density distribution have modelled the dark matter alone, while at the solar neighbourhood we expect the effects of the baryons, the gas and stars that make up the Milky Way, to be important. Read et al. \cite{Read:2008fh} have performed a series of simulations including the baryons, and have shown that massive satellites, dragged into the disc plane by dynamical friction, are torn apart by tidal forces depositing their stars and dark matter into a thick disc, lying in the same plane as the visible galaxy. Later the influence of this dark disc on direct detection of elastic dark matter was examined \cite{Bruch:2008rx}.  Here we investigate the influence of the dark disc on iDM detection.

Following \cite{Bruch:2008rx}, we assume the dark disc kinematics match the Milky Way's stellar thick disc, whose properties are listed in Table 1 of \cite{Read:2008fh}. This will be a good approximation if the Milky Way's stellar thick disc is mostly composed of accreted, rather than heated stars. We model the dark disc as a component of DM additional to the SHM, lagging the rotation of the Sun by $40$ km/s in the tangential direction, compared to the SHM which lags by $220$ km/s.  It is also assumed to have a Maxwellian velocity distribution, with a dispersion of ${\bf\sigma}=(63, 39, 39)$ km/s, and a density in the range $0.5 < \frac{\rho_{Disc}}{\rho_{SHM}} < 2$.  We employ these parameter ranges in our study of iDM signals from the dark disc.

Due to the smaller relative velocity of the dark disc with the Earth, when compared with the SHM, one would expect dark disc event rates to be low for high $\delta$ as the dark disc particles may not have high enough velocity to cross the minimum velocity threshold.

\section{Experimental information} \label{exper}

We now turn to a detailed discussion of the individual direct detection experiments.

\subsection{DAMA/LIBRA}

As the Earth revolves around the Sun, there should be a larger flux of WIMPs incident on the detector around the 2nd June, when the relative velocity of the Earth is at a maximum with respect to the galaxy. Conversely, the flux incident on the detector should be smallest around the 2nd December when the relative velocity of the Earth is at a minimum with respect to the galaxy. It is the annual modulation in the recoil event rate caused by this velocity modulation that the DAMA collaboration claim to have measured.

The first results from the DAMA/LIBRA set-up, with an exposure of 0.53 ton-yr, have recently been published \cite{Bernabei:2008yi}. These have been combined with the data collected by the DAMA/NaI set-up to give an impressive 0.82 ton-yr total exposure, yielding a modulation signal at 8.2 $\sigma$ C.L.

For a WIMP mass $M_{\chi} > 10\text{ GeV}$, this signal appears to be in conflict with other experiments under the assumptions of spin independent, elastic WIMP-nucleon scattering, with the WIMP phase space distribution described by the SHM. This has prompted a number of alternative explanations: light WIMPs \cite{Gelmini:2004gm, Gondolo:2005hh, Bottino:2005qj, Bottino:2008mf, Savage:2008er, Petriello:2008jj, Chang:2008xa,Andreas:2008xy}, spin dependent interactions \cite{Savage:2004fn}, mirror dark matter \cite{Foot:2003iv, Foot:2008nw} and iDM \cite{TuckerSmith:2001hy}.

\subsubsection{Quenching and channeling}

DAMA use highly radiopure NaI(Tl) scintillators as their target material. The light yield of scintillators depends on whether the recoiling nucleus interacts electromagnetically or via the strong nuclear interaction, since only electromagnetic interactions will produce photons. As a result, the measured energy, $E_{M}$, is different from the recoiling energy of the nucleus. This difference is expressed by the quenching factor $q$ defined by $E_{M}=q E_{R}$. We will follow the convention of measuring $E_{R}$ in keV, and $E_{M}$ in keVee (keV electron equivalent). The DAMA collaboration have measured the quenching factor for iodine and sodium for their NaI(Tl) crystals. They obtained the values $q_{Na}=0.3$ and $q_{I}=0.09$ \cite{Bernabei:1996vj} with an error of $0.01$ on the value of $q_{I}$ \cite{DAMA:quench}. Following the analysis of Ref.\cite{Chang:2008gd},
in our calculations, we use $q_{Na}=0.3$ and $q_{I}=0.085$, however in Section \ref{quench} we investigate what effect varying $q_{I}$ within the experimental limits has on our results.

If the nucleus recoils along certain directions in a crystalline structure, and if the recoiling energy is low, no nuclear interactions occur, so the quenching factor, q, equals one. This effect is known as channeling \cite{Bernabei:2007hw}. Since NaI(Tl) is a crystalline material, we need to include this effect in our calculations. We use the parameterisation given in \cite{Fairbairn:2008gz} for the fraction $f$ of channeled events relevant for DAMA
\begin{equation} f_{Na}(E_{R})\approx \frac{e^{-E_{R}/18}}{1+0.75 E_{R}}, \qquad f_{I}(E_{R})\approx \frac{e^{-E_{R}/40}}{1+0.65 E_{R}}.
\end{equation}

\subsubsection{Calculating limits}

DAMA have released their binned data from the combined DAMA/LIBRA and DAMA/NaI data sets covering the range 2-20 keVee. A clear modulation signal is present below about 8 keVee, while the modulation is consistent with zero at higher energies \cite{Bernabei:2008yi}. The rate in the lowest energy bin, covering the range 2-2.5 keVee, is smaller than in the next bin, covering the range 2.5-3 keVee, suggesting that the rate is falling to zero at low energies. While care should always be taken when examining the data at the edges of the experimental sensitivity, it should be noted that both of these features are found naturally with inelastic scattering.

We use a $\chi^2$ goodness of fit test to analyse the DAMA results. We construct a $\chi^2$ function using the twelve 0.5 keVee width bins between 2.0 - 8.0 keVee and their relative uncertainties. We do not fit to the higher energy bins because, as mentioned above, the inelastic spectrum falls off at high energies.

In our 2D plots, we find the best fit point by minimising the $\chi^2$ function for the two unconstrained parameters, either $\sigma_n$ and $M_{\chi}$, or $\sigma_n$ and $\delta$. For a goodness of fit, we require that the best fit point have $\chi^2_{min}<10$, given that we have 12 bins and 2 free parameters. Allowed regions at a given confidence limit are obtained by looking for contours $\chi^2=\chi^2_{min}+\Delta \chi^2$, where $\Delta \chi^2 = 4.61$ or $10.60$ for 90\% and 99.5\% confidence limits respectively. 

DAMA have also released the unmodulated rate for a single-hit scintillation as measured by DAMA/LIBRA. A WIMP is expected to scatter once, so we are using the rest of the detector as a veto. No other background subtraction is applied. We can use this data to set a limit by requiring that the calculated unmodulated rate not be larger than the measured rate across the energy range shown in Figure 1 of \cite{Bernabei:2008yi}.

It should be noted that we have not included the effect of the finite energy resolution of the detector in our calculations. We are fitting to the modulated rate from the combined data sets of DAMA/LIBRA and DAMA/NaI, however the energy resolution for DAMA/LIBRA and DAMA/NaI are known to be different \cite{Bernabei:2008yh, Bernabei:1998rv}.

\subsection{Null experiments}
In this section, we discuss the data used in calculating the regions excluded at 90\% confidence level by other direct detection experiments. We adopt a conservative approach and include events that the experiments ascribe to background processes when setting limits. We will find that the most constraining experiment is CRESST-II. Unless otherwise stated, we use the ``pmax method", described in \cite{yellin}, in setting our limits. This method has the advantage of setting similarly strong limits as the ``optimum interval method", but is easier to implement. Care should be taken when comparing exclusion curves at 90\% confidence level and the region allowed by DAMA at 90\% confidence limit, since we are using different statistical methods in evaluating them. In calculating the total event rate, we integrate the differential rate over the energy range of the experiment, then find the average value over the running dates given by the collaboration, and finally, we multiply by the exposure.  As a final check of our methods, we have reproduced  the published exclusion curves from CDMS II and XENON10 in the  $\delta=0$ limit.

\subsubsection{CDMS II}
In setting the limits for CDMS II, we use the three runs from the Soudan Underground Laboratory which were sensitive to nuclear recoil energies between 10-100 keV \cite{Akerib:2004fq, Akerib:2005kh, Ahmed:2008eu}. We only consider scattering from germanium as scattering off of silicon is highly suppressed. The published effective germanium exposures are weighted for a WIMP mass of 60 GeV and averaged over recoil energies 10-100 keV, however these numbers are expected to apply for an inelastic WIMP since their acceptance efficiency is fairly constant over their energy range. The first run took place from 11th October 2003 to 11th January 2004, had an exposure of 19.4 kg-day and saw one event at 64 keV \cite{Akerib:2004fq}. The second run was from 25th March 2004 to 8th August 2004, had an exposure of 34 kg-day and saw one event at 10.5 keV \cite{Akerib:2005kh}. The latest five-tower run was between October 2006 and July 2007, had an exposure of 121.3 kg-day and saw no events \cite{Ahmed:2008eu}.

\subsubsection{CRESST-II}
For the CRESST-II limits, we use data collected by the Julia and Daisy detectors \cite{Angloher:2004tr} between 31st January 2004 - 23rd March 2004, and the Verena and Zora detectors \cite{Angloher:2008jj} between 27th March 2007 - 23rd July 2007. We only consider scattering off of the tungsten atoms in the CaWO$_4$ crystals, and include events on or beneath the curve where 90\% of the tungsten recoils are expected. The Julia detector had a tungsten exposure of 6.26 kg-day and we include the four observed events between 10 - 50 keV and an unpublished event above 50 keV \cite{HKrauss}. The Daisy detector had a tungsten exposure of 6.84 kg-day and we include the two events between 10-12 keV, the event which lies on the curve where 90\% of the tungsten recoils are expected, at approximately 22 keV, the event at approximately 45 keV, and a second unpublished event above 50 keV \cite{HKrauss}. The Verena and Zora detectors had a combined tungsten exposure of 30.6 kg-day and we use the seven published events between 10 - 100 keV. To take into account the difference in energy regions, we do not integrate above the energy of the extra events for the Daisy and Julia detectors. 

Note that following \cite{Chang:2008gd} we have set limits by including the Daisy point which lies on the tungsten curve at 22 keV in Figure 9 of Ref.\cite{Angloher:2004tr}.  If we redo the analysis by ignoring this point and only including data which lie fully below the curve, we find that CRESST-II just manages to exclude all of the DAMA/LIBRA 99.5\% allowed region. This point has such a large impact on the limits set using the ``pmax method" because it lies at the peak of the recoil energy spectrum, as can be seen in the left panel of Figure \ref{fig:formplot}.

This shows us that with further data, the CRESST-II experiment has the ability to completely exclude the DAMA/LIBRA region.  

\subsubsection{KIMS}
From the null experiments we consider, KIMS \cite{Lee.:2007qn} is the only one where scattering occurs off of iodine, in the CsI(Tl) crystals they employ. Two of the crystals collected experimental data between June 2005 and March 2006, then another two crystals were installed, running between December 2005 and March 2006 \cite{HSLee}. In setting limits, we find the average rate and statistical error from the four crystals, and require the calculated rate to be less than the measured rate plus 1.64 times the error in the first five bins, corresponding to the energy range 3-8 keVee. For the quenching factor, we use the parameterization
\begin{equation} q_{CsI(Tl)} (E_{R})\approx \frac{0.175 e^{-E_{R}/137}}{1+0.00091 E_{R}}.
\end{equation}
This function fits the curve shown in Figure 13 of \cite{Park:2002jr} to a good accuracy.

\subsubsection{XENON10}
XENON10 \cite{Angle:2007uj} is a liquid xenon based experiment which ran from 6th October 2006 until 14th February 2007 and had an effective exposure of 316.4 kg-day. In calculating our limits, we include the 10 events used in their analysis in the range 4.5 - 26.9 keV, as well as the 14 events in the range 26.9 - 45 keV. Below 26.9 keV, we use the published acceptances and above 26.9 keV we use the known software cut efficiencies and assume a constant nuclear recoil acceptance of 0.45 \cite{Kni}.

\subsubsection{ZEPLIN-II}
We include only the limits from ZEPLIN-II \cite{Alner:2007ja} since ZEPLIN-I \cite{Alner:2005pa} is not competitive with the other experiments we consider \cite{Chang:2008gd}. ZEPLIN-II is another xenon based experiment with an effective exposure of 225 kg-day and was assumed to run between May and July 2006, as suggested by \cite{ZIIrun}. 29 events were observed in the range 5 - 20 keVee, which corresponds to a range of 13.9 - 55.6 keV if the published quenching factor $q=0.36$ is used. In calculating our limits, we use the published efficiencies and we take all observed events as signal.

\subsubsection{ZEPLIN-III}

We also include the first results from the ZEPLIN-III experiment \cite{Lebedenko:2008gb}, the successor to ZEPLIN-II. ZEPLIN-III is also xenon based and improves on the limits set by ZEPLIN-II as only 7 events were observed, in the range 10.7 - 30.2 keV. The effective exposure was 126.7 kg-day, the data being collected between 27th February 2008 and 20th May 2008. Using Figure 15 of \cite{Lebedenko:2008gb}, we extract an approximate quenching factor parameterization:
\begin{equation}
q_{Xe}(E_{M}) \approx (0.142 E_{M} + 0.005) \exp[-0.305 (E_{M})^{0.564}].
\end{equation}
We extracted the energies of the 7 observed events from Figure 12 of \cite{Lebedenko:2008gb} and converted to nuclear recoil energies using the above quenching factor.

\section{Results and discussion} \label{Results}
In this section we present our results for iDM in the context of the different halo models.

\subsection{$\text{VL}_{220}$ vs SHM}
\begin{figure}[h!t]
\centering
\includegraphics[height=2.85in,width=0.5\textwidth]{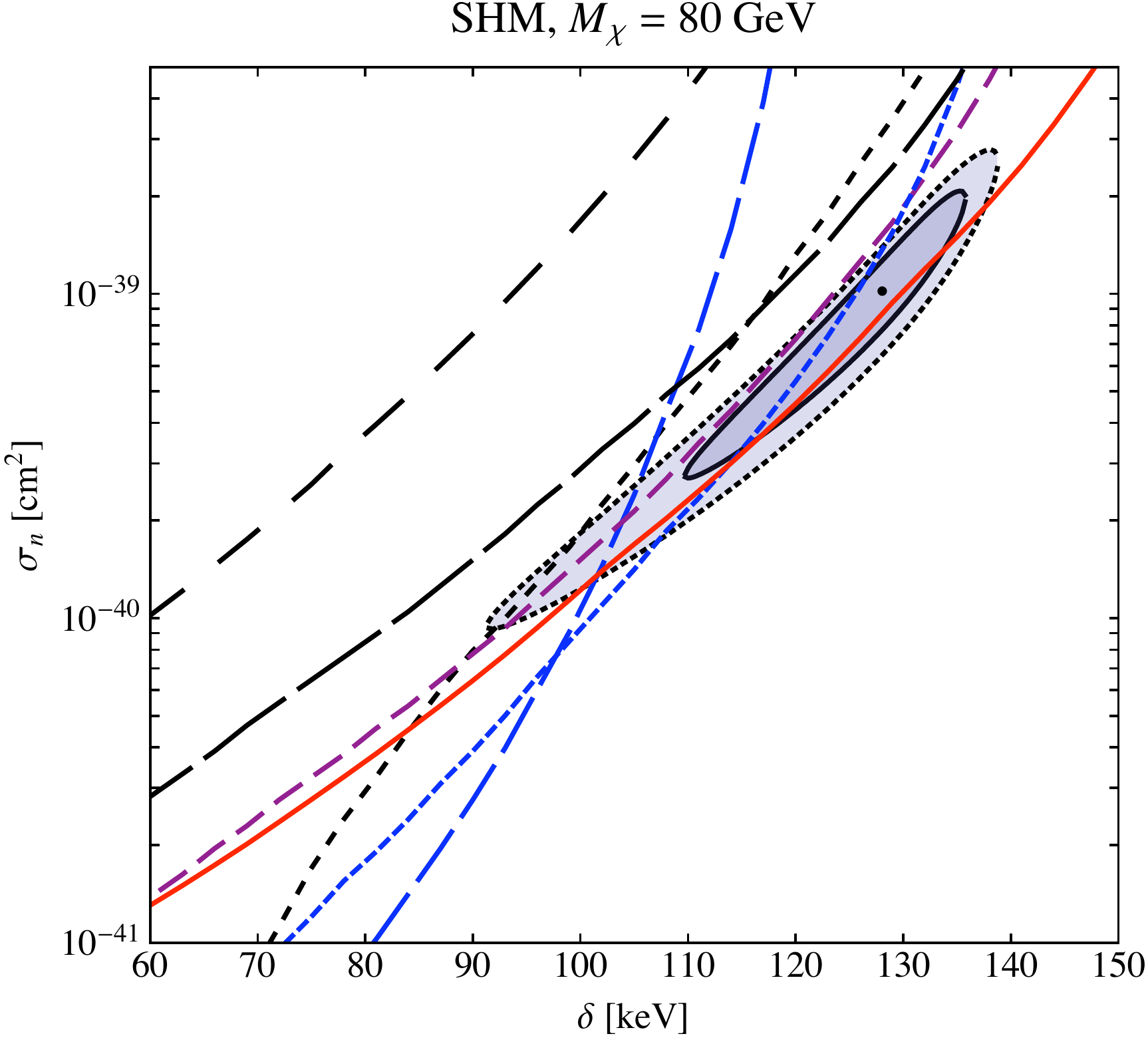}\includegraphics[height=2.85in,width=0.5\textwidth]{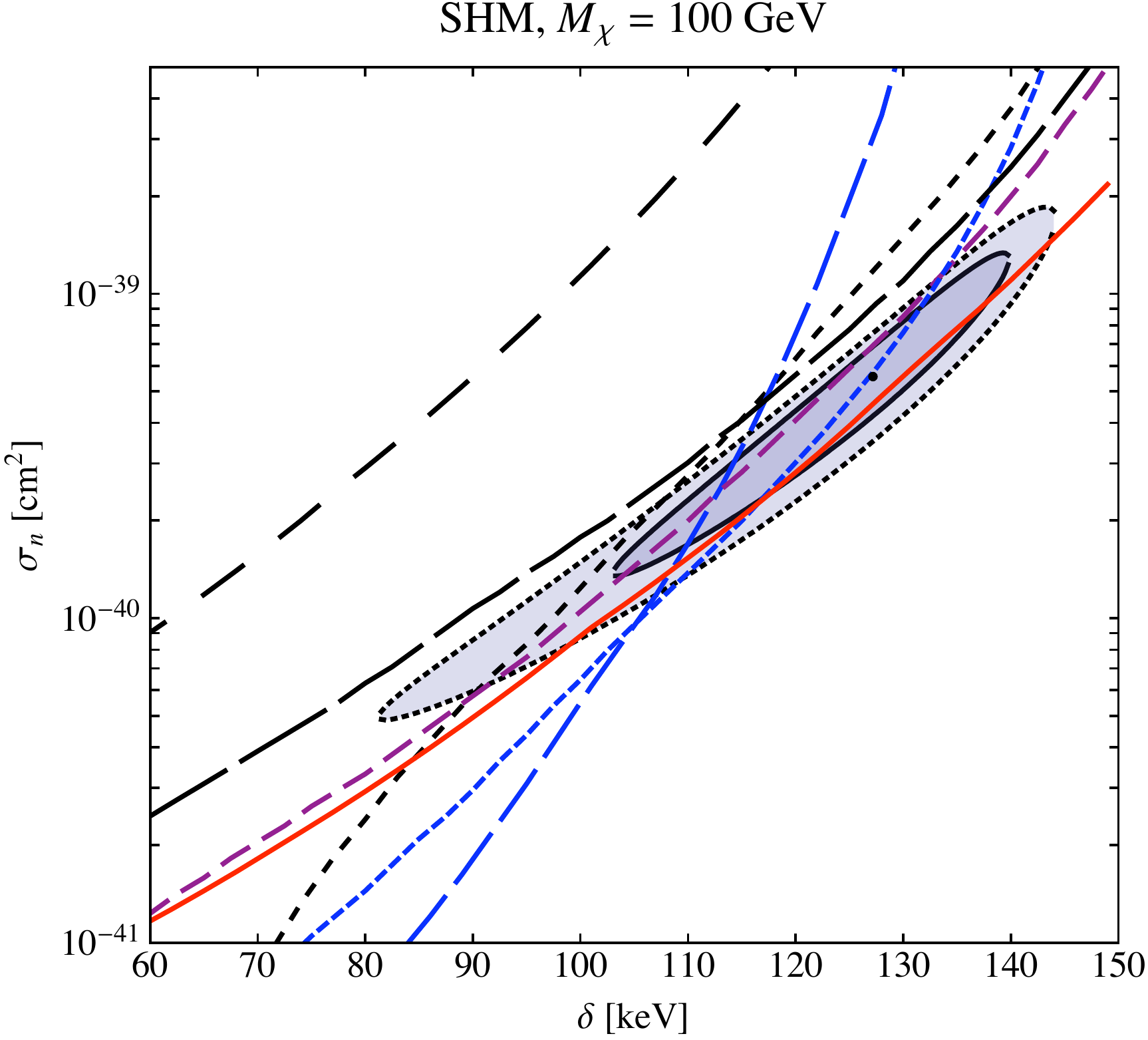}
\includegraphics[height=2.85in,width=0.5\textwidth]{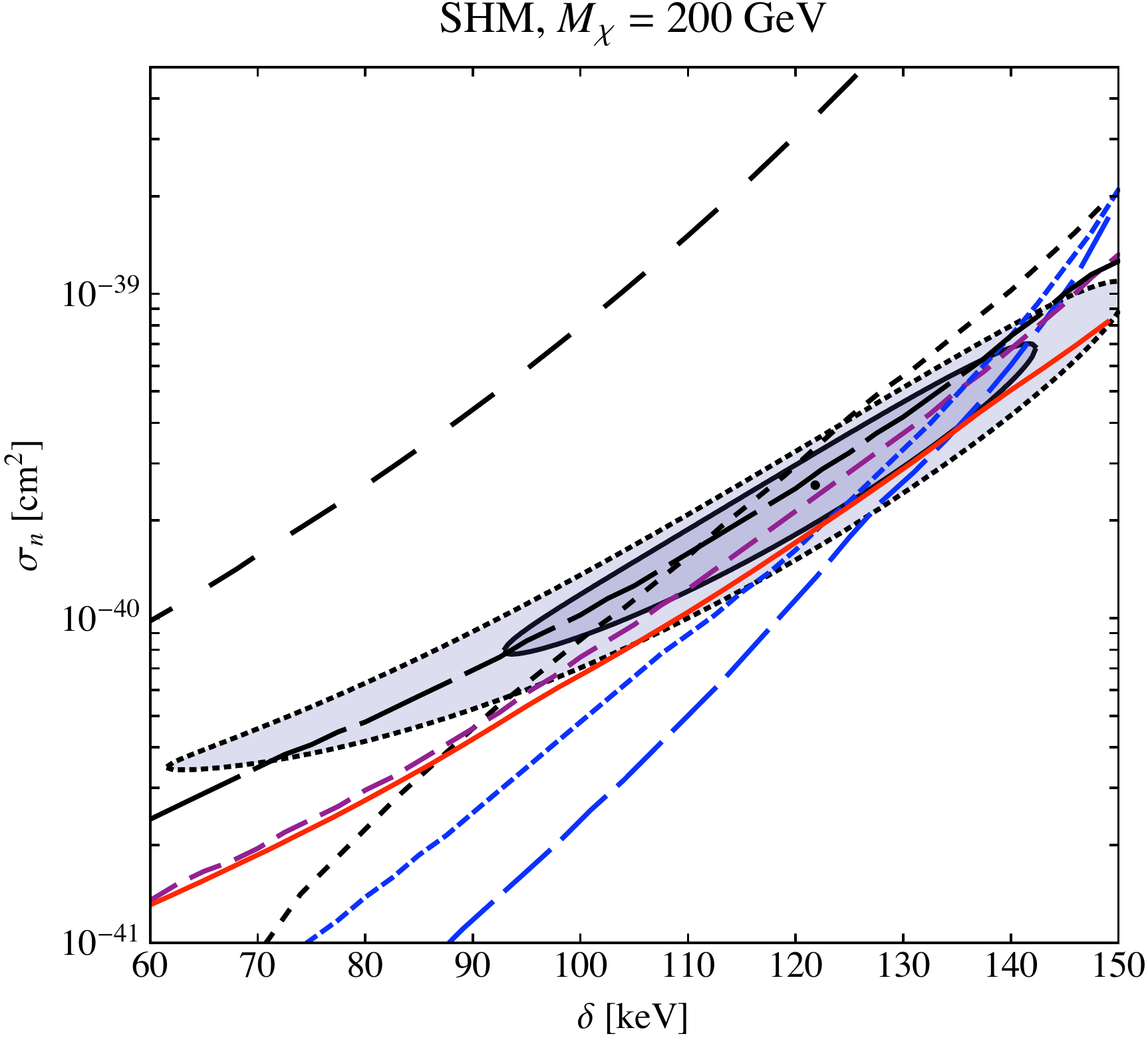}\includegraphics[height=2.85in,width=0.5\textwidth]{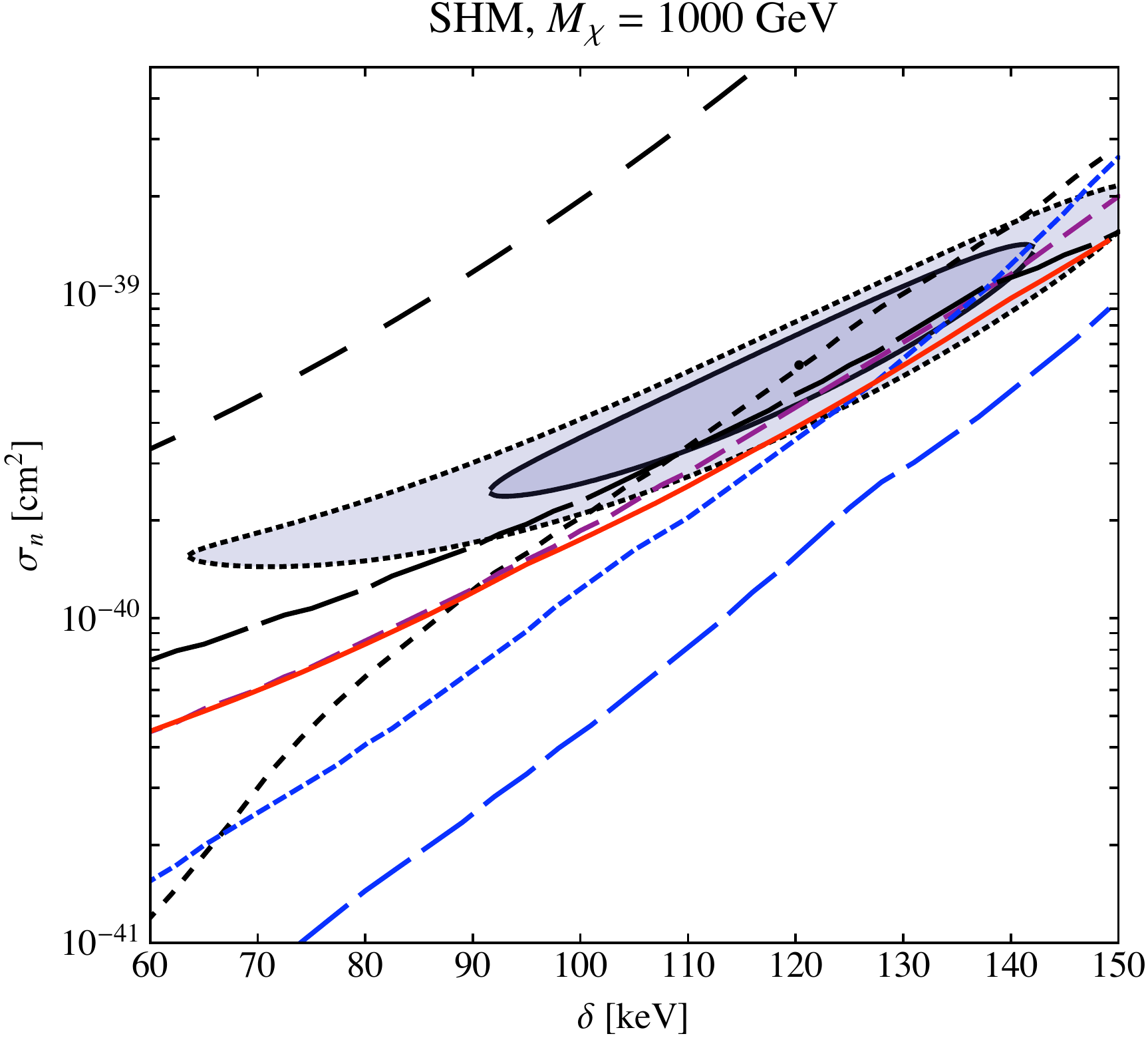}\\
\includegraphics[height=1.5in]{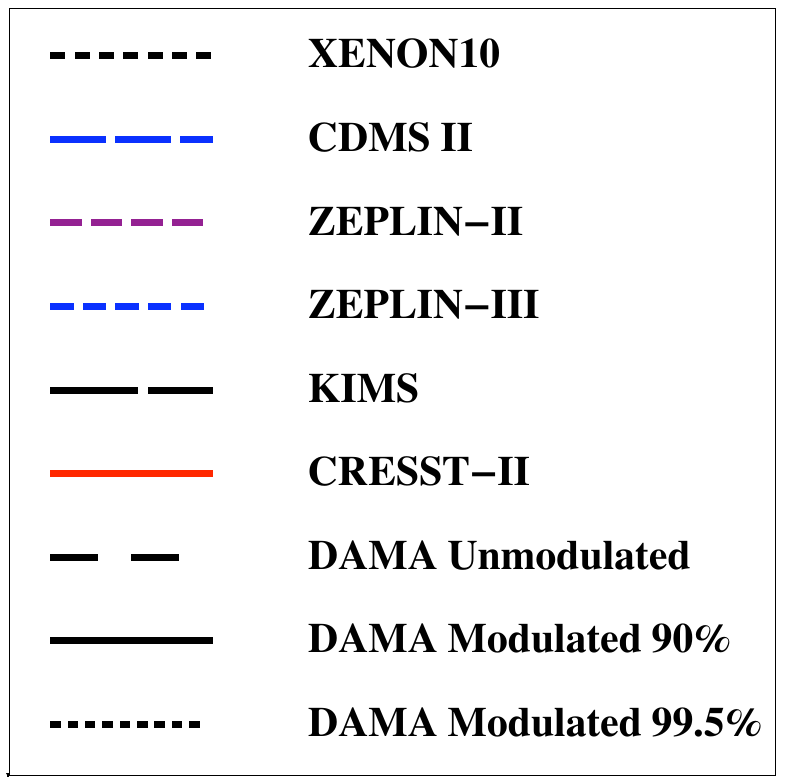}
\caption{Here we show the variation in the exclusion limits set by the experiments as $\delta$ is varied and $M_{\chi}$ is held constant. These limits are calculated using the SHM. The preferred region of parameter space for the DAMA results is shown at 90\% and 99.5\%, and the DAMA best fit point is plotted with a dot. As one can see there is a small region of agreement between all experiments for low masses and $\delta \sim 130$ keV.  At higher masses both CRESST-II and CDMS II exclude the DAMA results and the region of agreement with the other experiments is greatly reduced.}\label{deltaplots}
\end{figure}

\begin{figure}[h]
\centering
\includegraphics[height=2.85in,width=0.5\textwidth]{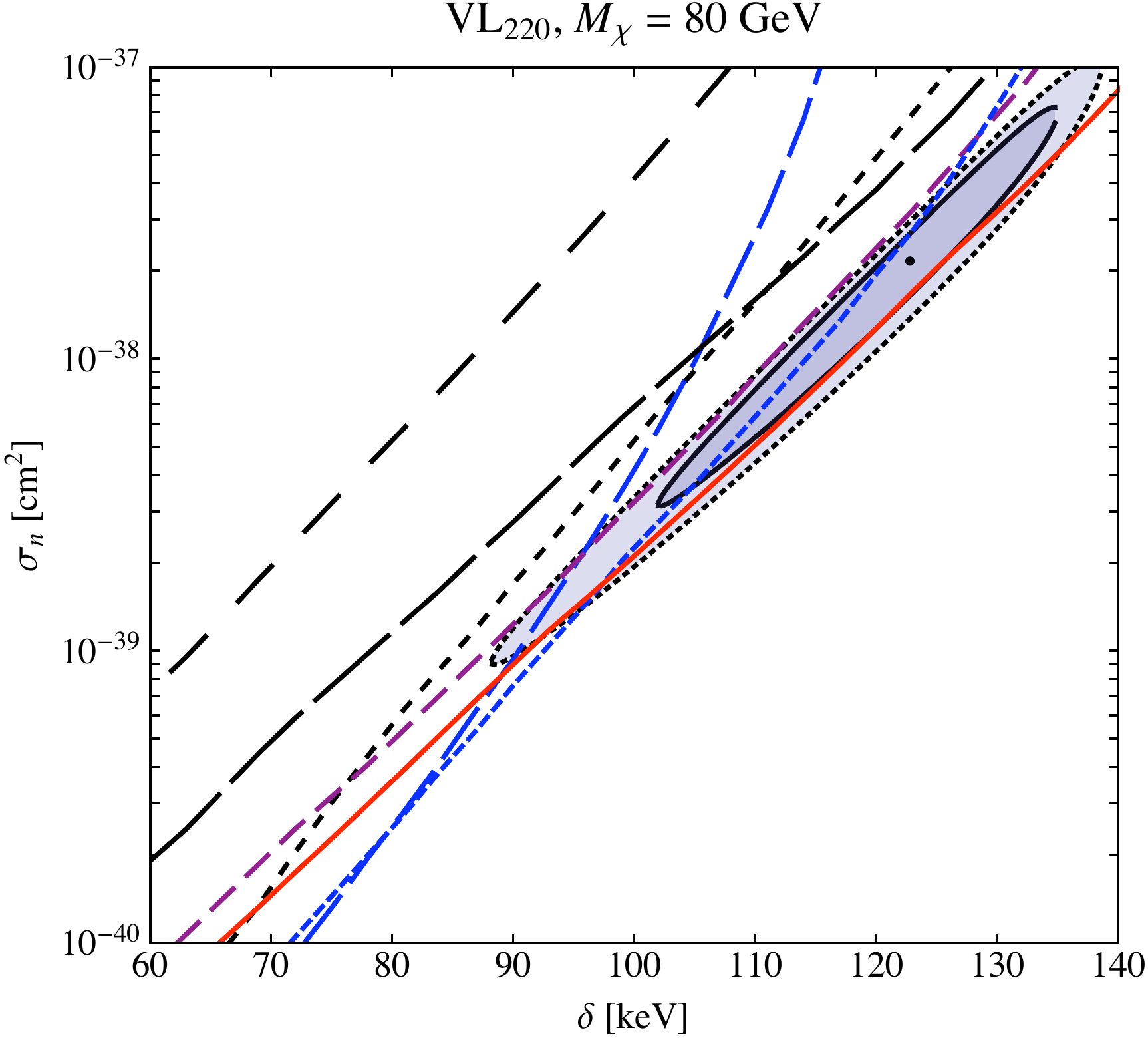}\includegraphics[height=2.85in,width=0.5\textwidth]{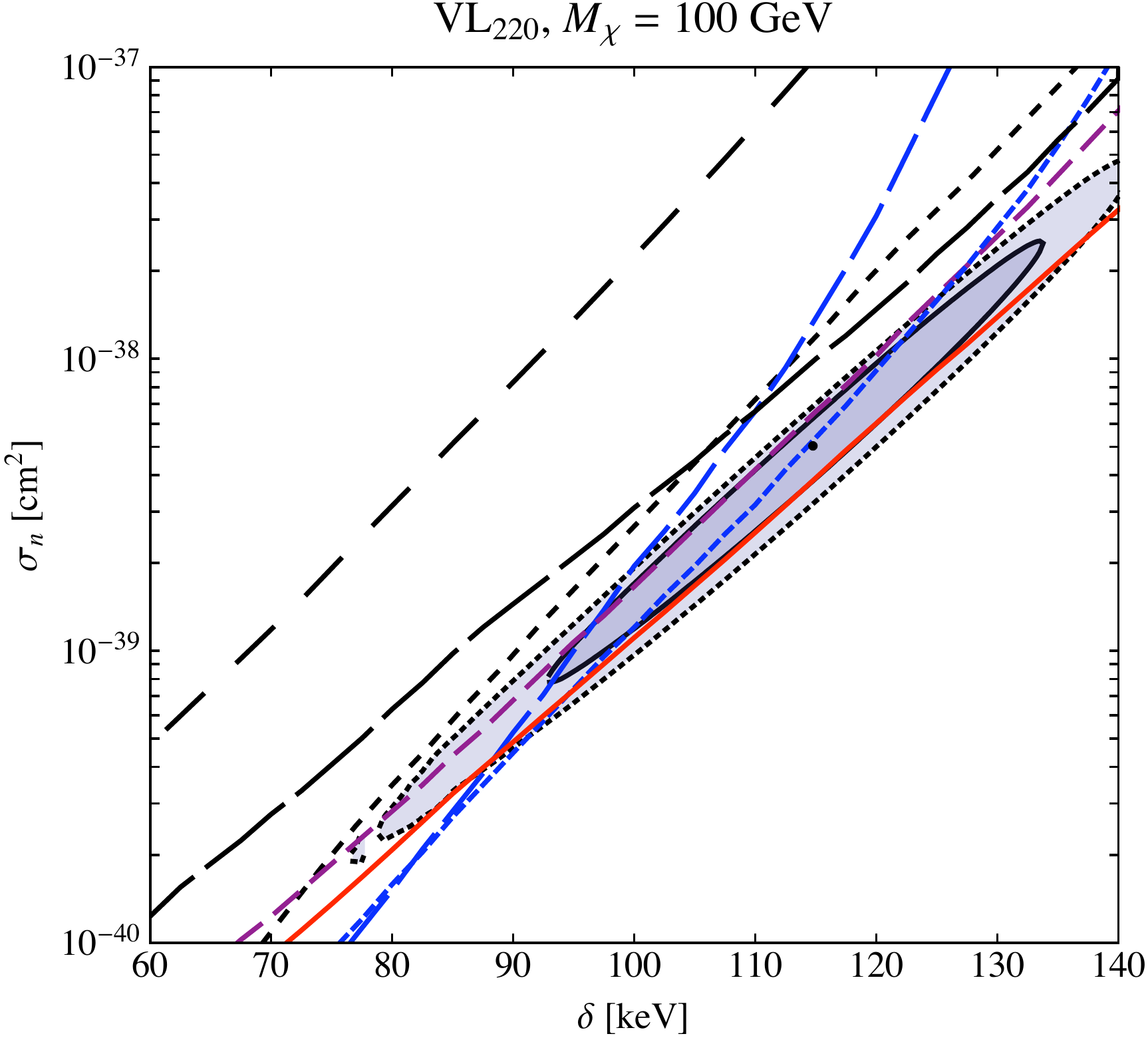}
\includegraphics[height=2.85in,width=0.5\textwidth]{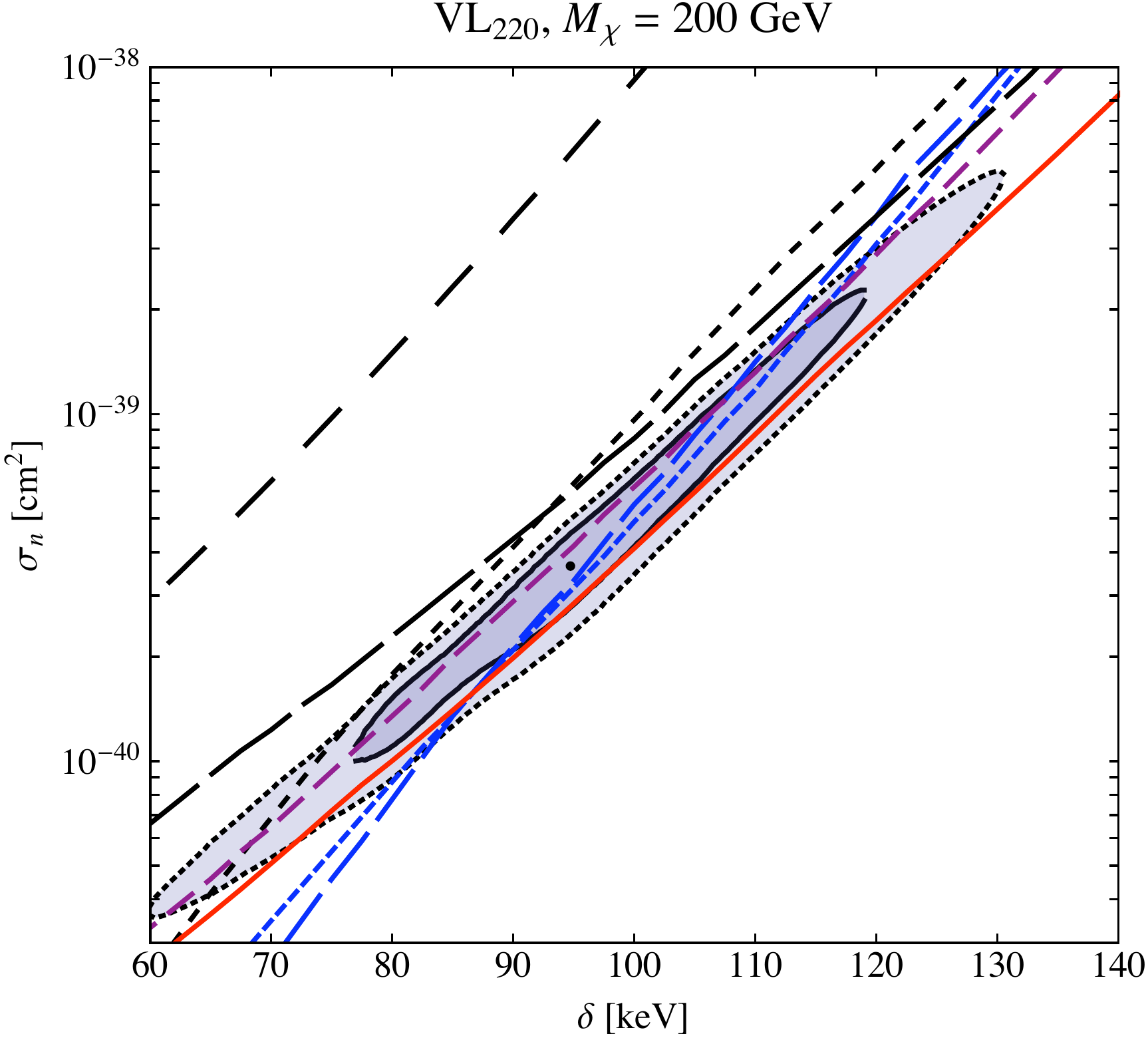}\includegraphics[height=2.85in,width=0.5\textwidth]{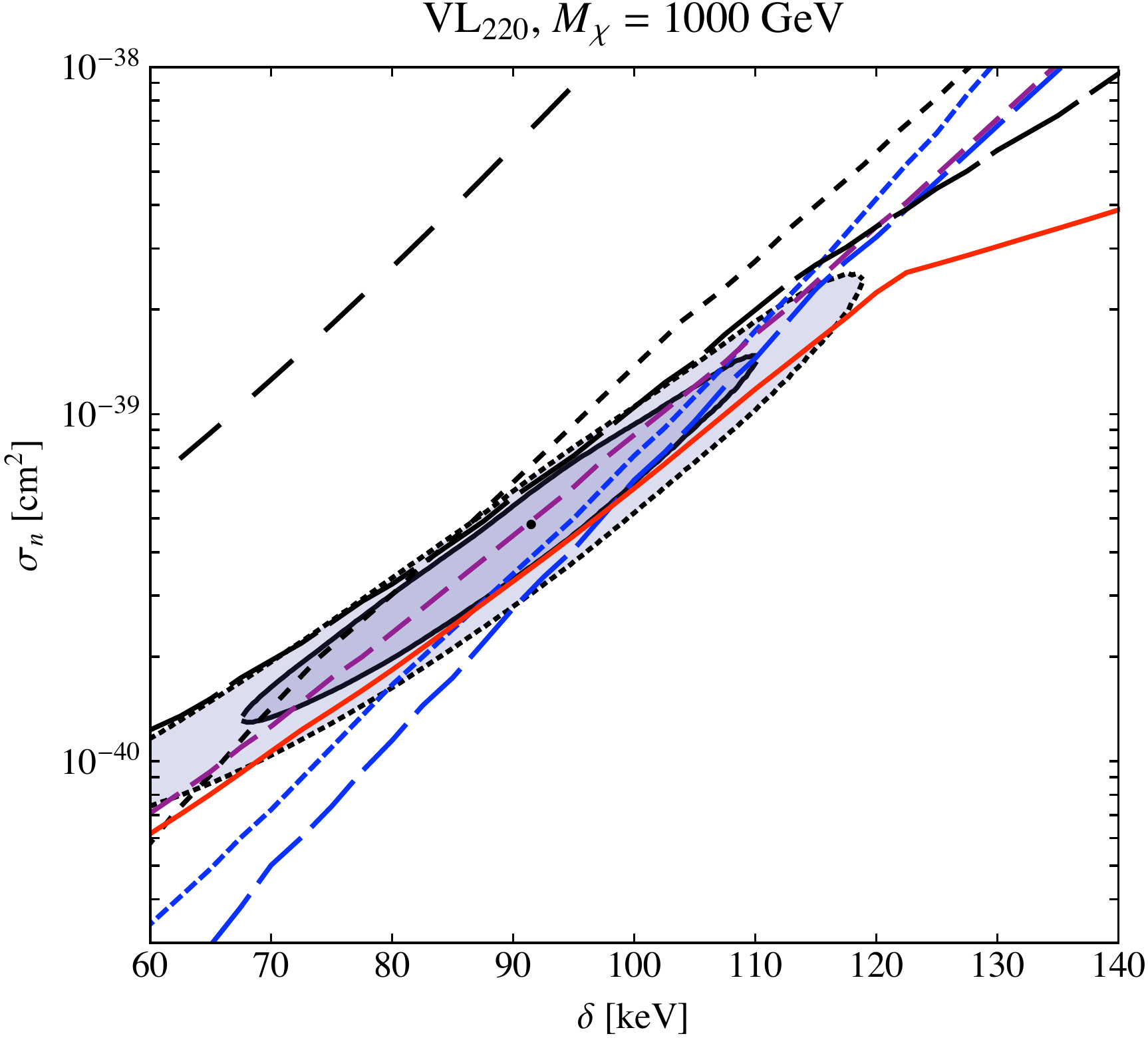}\\
\caption{Change in limits when the SHM is replaced by the $\text{VL}_{220}$ halo, cf, Figure 2. At low masses there is a smaller region of agreement between CRESST-II and DAMA, with CRESST-II almost excluding DAMA at the 90\% level over all masses. At high masses, however, CRESST-II and CDMS II are significantly less constraining on the DAMA region than for the SHM, this can be seen by comparing the bottom right panels of both figures. One can also see that the typical cross sections are an order of magnitude higher for the $\text{VL}_{220}$ halo at low masses than for the SHM (note also that the lower panels have a different scale for the cross section).}\label{deltaplots2}
\end{figure}

In Figures \ref{deltaplots} and \ref{deltaplots2} we present a comparison of the iDM limits on the DAMA/LIBRA preferred region under the assumptions of the SHM and the $\text{VL}_{220}$ halo, for a variety of WIMP masses.

As one can see the allowed cross section limits are generally higher for the $\text{VL}_{220}$ halo than the SHM. Also the CRESST-II results are slightly more constraining on the DAMA/LIBRA preferred region at low mass when the $\text{VL}_{220}$ halo is used. The most interesting feature to note is that all of the experiments show less disagreement with the DAMA/LIBRA region at high WIMP masses when the $\text{VL}_{220}$ halo is used. This feature is most prominent for the CDMS II exclusion line, but is true for all experiments to some degree.

\begin{figure}[h]
\centering
\includegraphics[height=2.8in,width=0.5\textwidth]{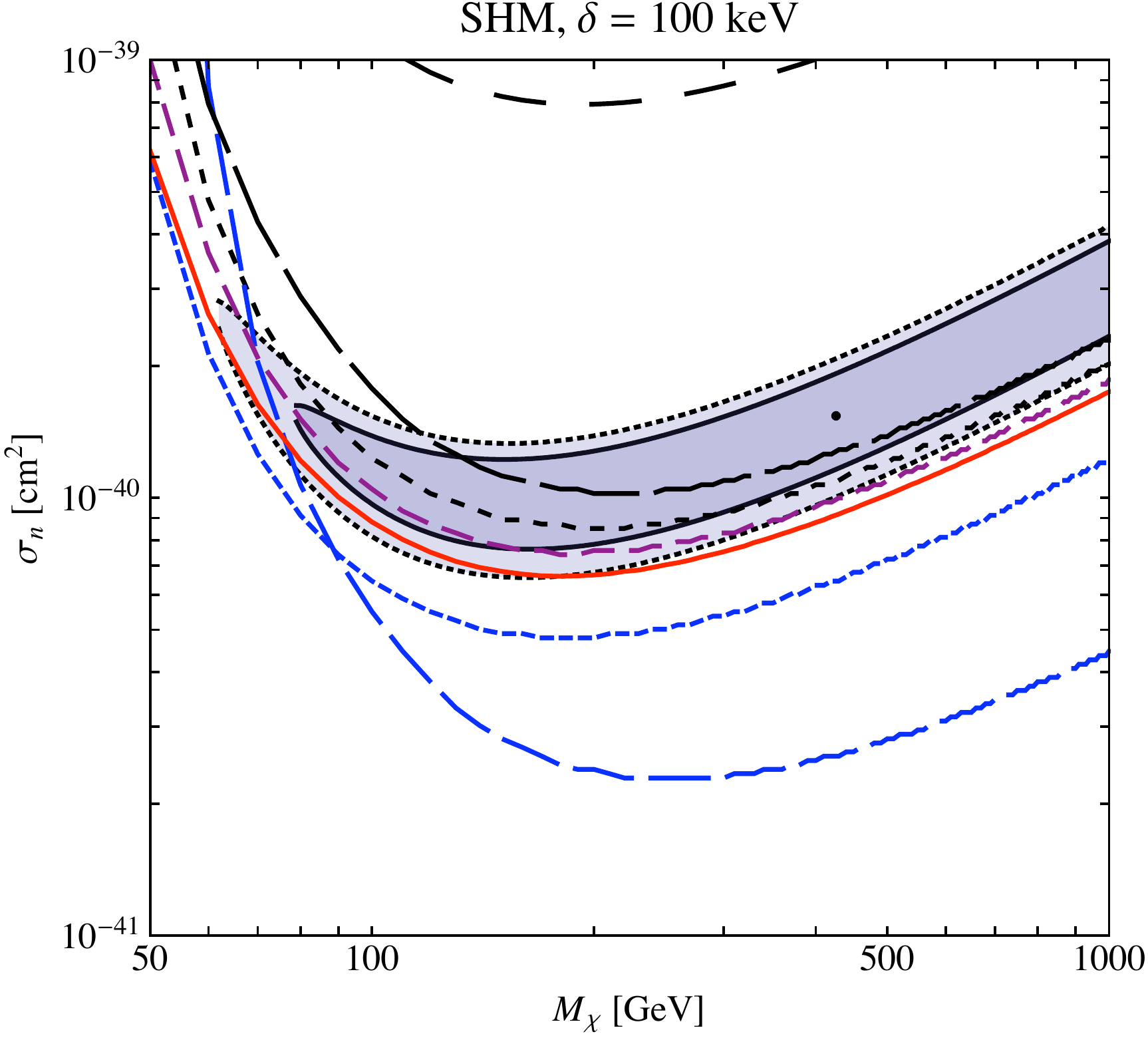}\includegraphics[height=2.8in,width=0.5\textwidth]{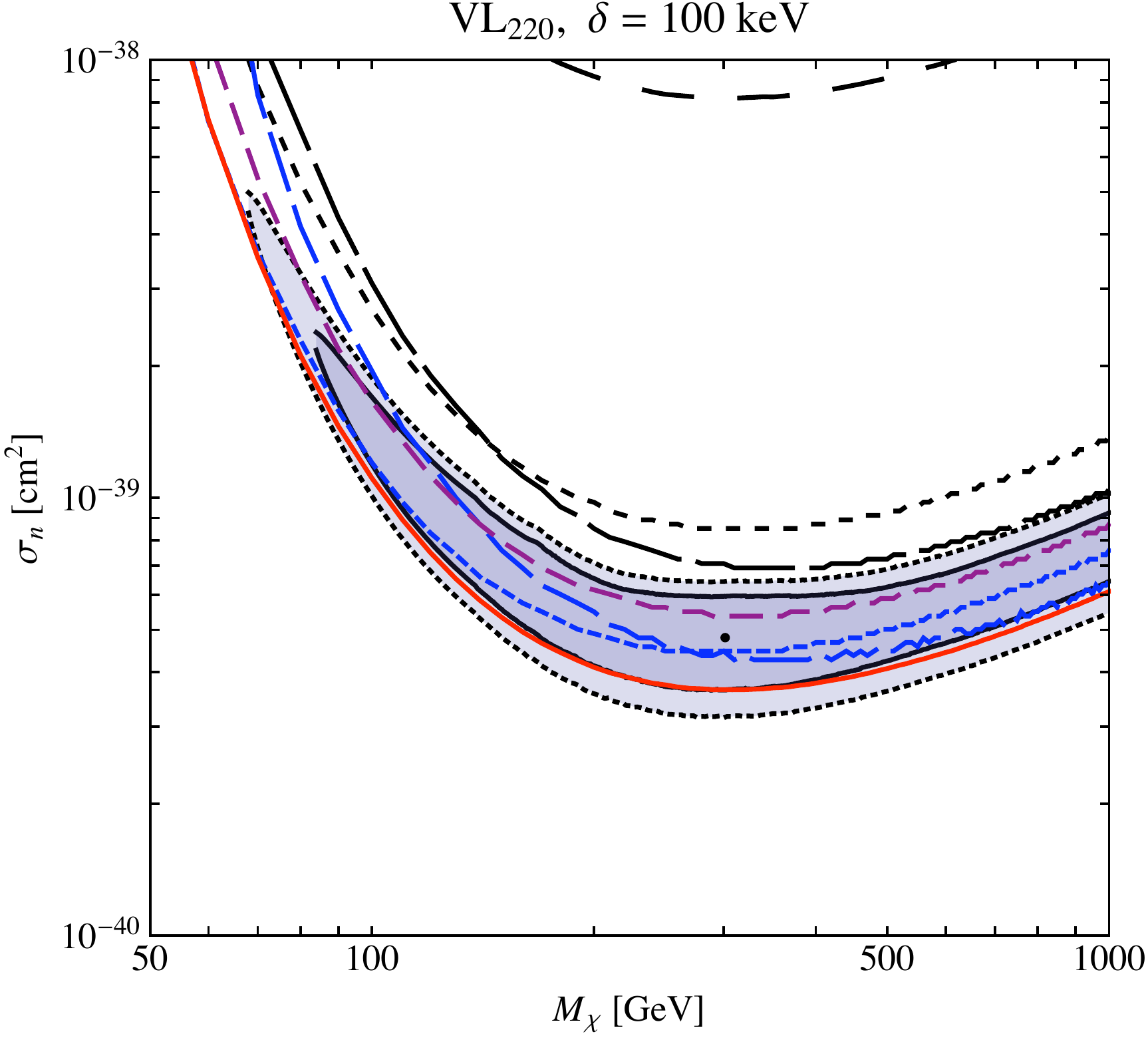}
\caption{The allowed parameter space for fixed $\delta=100$keV and varying $M_{\chi}$. For the SHM (left panel) and this value of $\delta$, CDMS II excludes the DAMA region at 90\%.  For the $\text{VL}_{220}$ halo (right panel) the tightest constraints are set by CRESST-II, and there is agreement between DAMA and CDMS II up to high WIMP masses. Again one can see that the typical allowed cross sections are an order of magnitude higher for the $\text{VL}_{220}$ halo.}\label{massplots}
\end{figure}

\begin{figure}[h]
\centering
\includegraphics[height=2.6in]{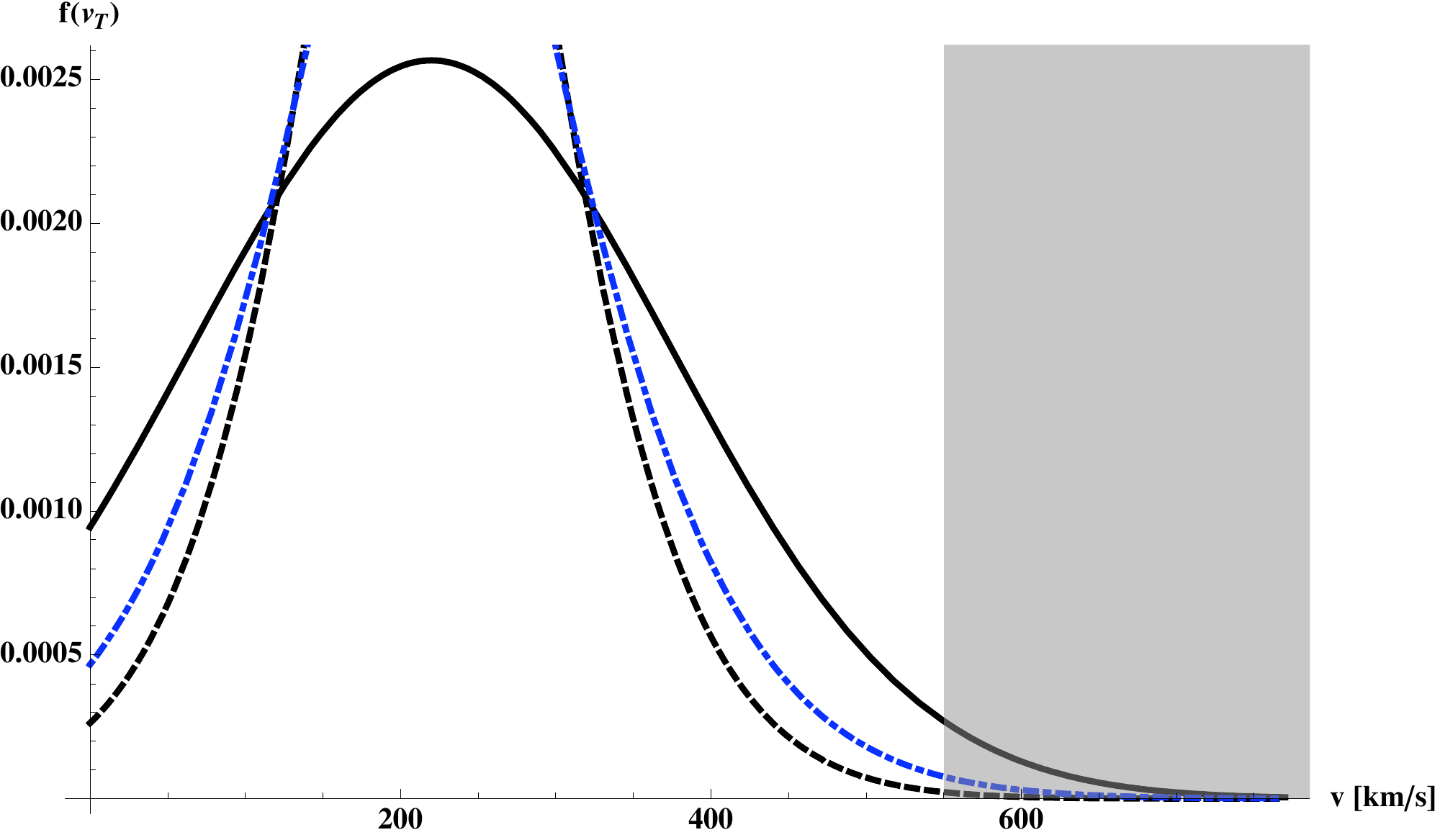}\caption{Detectable particle distributions in a germanium detector as a function of tangential velocity.  The detectable region is shaded in grey. The solid, dashed and dot-dashed lines show the SHM, $\text{VL}_{220}$ and $\text{VL}_{270}$ distributions respectively.  The left edge of the grey region corresponds to $M_{\chi}=500 \text{ GeV}$.}\label{CDMSVEL}
\end{figure}

This feature is made more explicit by observing how the limits change as a function of $M_\chi$ for fixed $\delta$.  In Figure \ref{massplots} we show the exclusion limits for $\delta = 100$ keV and $50 \text{ GeV} < M_\chi < 1000 \text{ GeV}$.

In this case all experiments are less constraining on DAMA/LIBRA at high mass, however the CDMS II limits completely rule out the entire DAMA/LIBRA preferred region under the assumptions of the SHM, yet for the $\text{VL}_{220}$ halo there is agreement up to $M_\chi \sim O(\text{ TeV})$.  This can be explained by a combined effect of iDM and the $\text{VL}_{220}$ velocity distribution, as noted in Section \ref{haloes}.

Looking quantitatively at this effect we can take the example of germanium in the CDMS II detector.  As the inelasticity has pushed the minimum velocity up significantly we need only consider the tangential velocity distributions (in the Earth's frame), as they are centered around $v_{circ} \sim 220 \text{ km/s}$ whereas the radial velocity distribution is centered around ${v_0}_r \sim 0 \text{ km/s}$.  Therefore the high velocity components scattering on the germanium will be from the tangential component of the halos.

The CDMS II detector is sensitive to the energy range $10 \text{ keV} < E_R < 100 \text{ keV}$, and if we consider $50 \text{ GeV} < M_\chi < 500 \text{ GeV}$, then for $\delta = 100$ keV the minimum velocity lies in the range $550 \text{ km/s} < v_{min} < 1006 \text{ km/s}$.  As the escape velocity is taken as $v_{esc} = 550 \text{km/s}$ in the halo rest frame, then the highest velocity particles observed in the Earth's frame have $v_{max} \sim 770 \text{km/s}$.  Therefore the observable particles will have velocities $550 \text{ km/s} < v_{obs} < 770 \text{ km/s}$.  In Figure \ref{CDMSVEL} we plot the two different tangential velocity distributions and the region over which $v_{min}$ varies as $M_{\chi}$ is increased.

As one can see the detectable particle number in the SHM distribution grows significantly as $M_{\chi}$ is increased and $v_{min}$ decreases, leading to stronger limits on the cross section from the CDMS II results.  However, for the $\text{VL}_{220}$ distribution this growth in the number of detectable particles is not nearly as pronounced, due to the reduced width of the distribution, and this leads to much slower variation in CDMS II exclusion limits as $M_{\chi}$ is varied, and overall to much weaker limits than from the SHM.  As stated above this effect is due to the combination of iDM, which increases the minimum velocity relative to elastic scattering, and the $\text{VL}_{220}$ velocity distribution, which is much smaller than the SHM at high velocities.

\subsection{$\text{VL}_{270}$ vs SHM}
\begin{figure}[h]
\centering
\includegraphics[height=2.85in,width=0.5\textwidth]{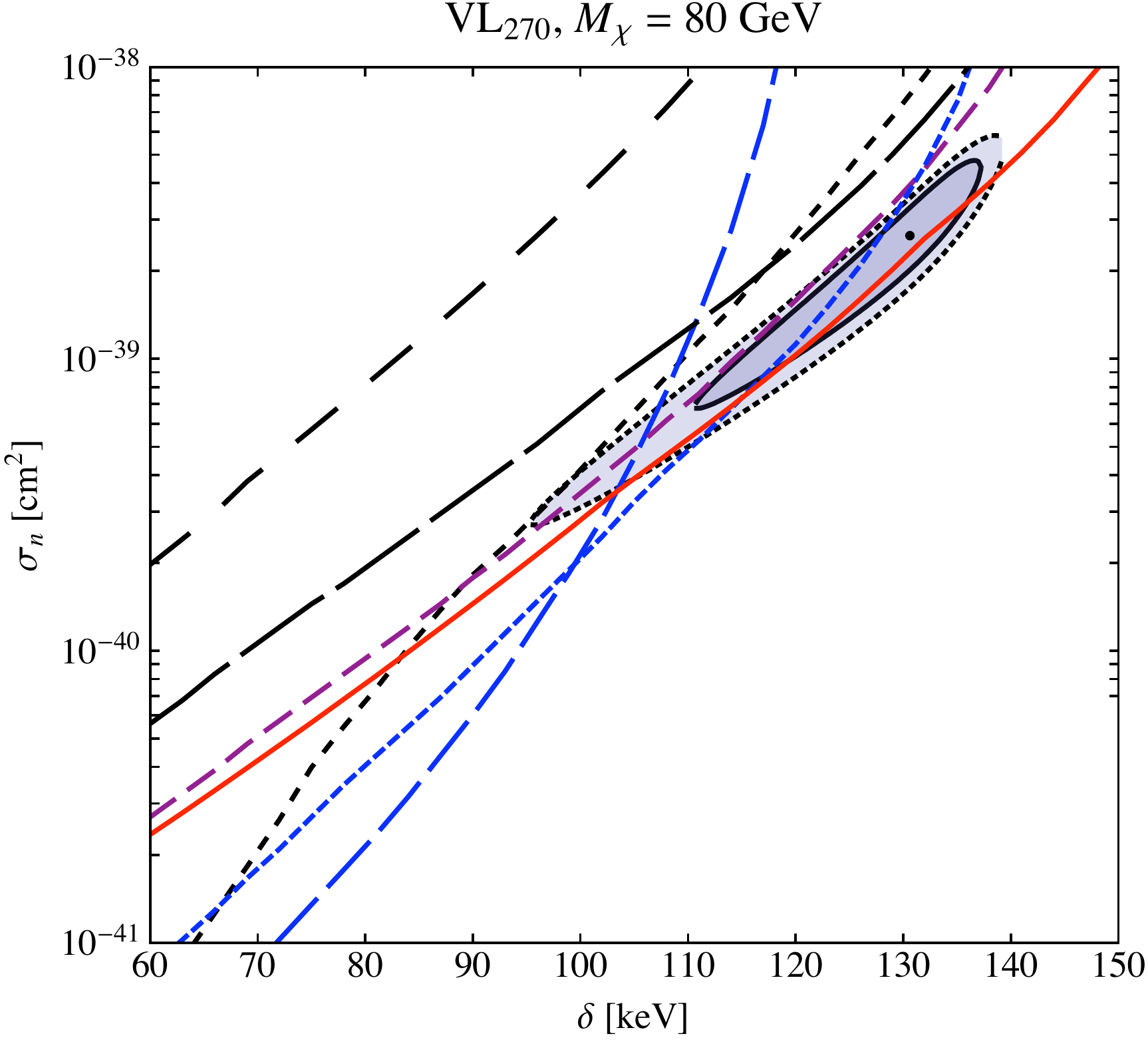}\includegraphics[height=2.85in,width=0.5\textwidth]{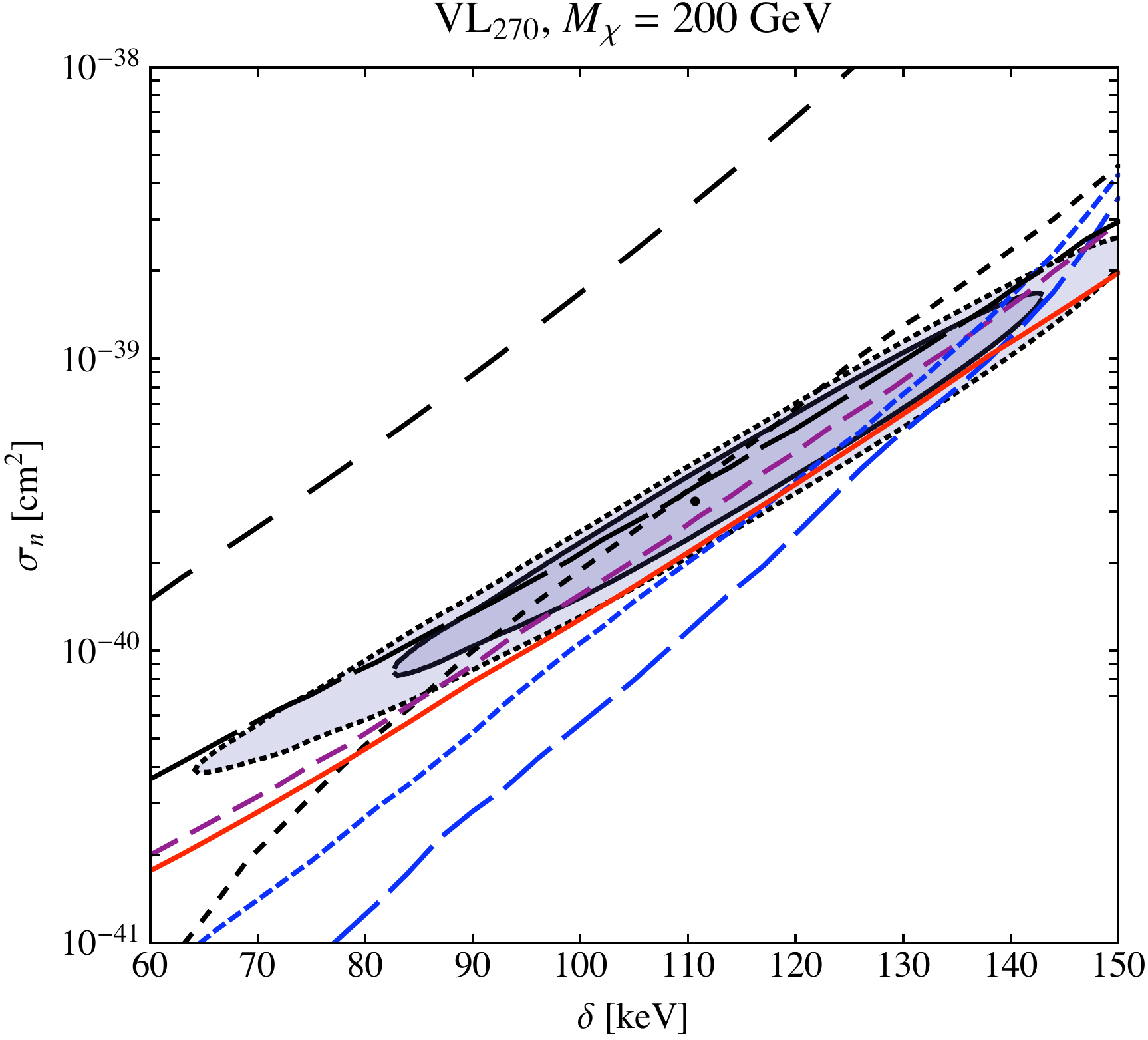}\\
\caption{Exclusion limits for $\text{VL}_{270}$. These results are similar to those presented in Figure 2. We see that there are still allowed regions of the DAMA parameter space at the 90\% confidence level for WIMP masses lower than $200$ GeV and $\delta \sim 130$ keV, however these typically occur at larger cross sections than in Figure 2.}\label{VL270deltaplots}
\end{figure}

\begin{figure}[h]
\centering
\includegraphics[height=2.8in,width=0.5\textwidth]{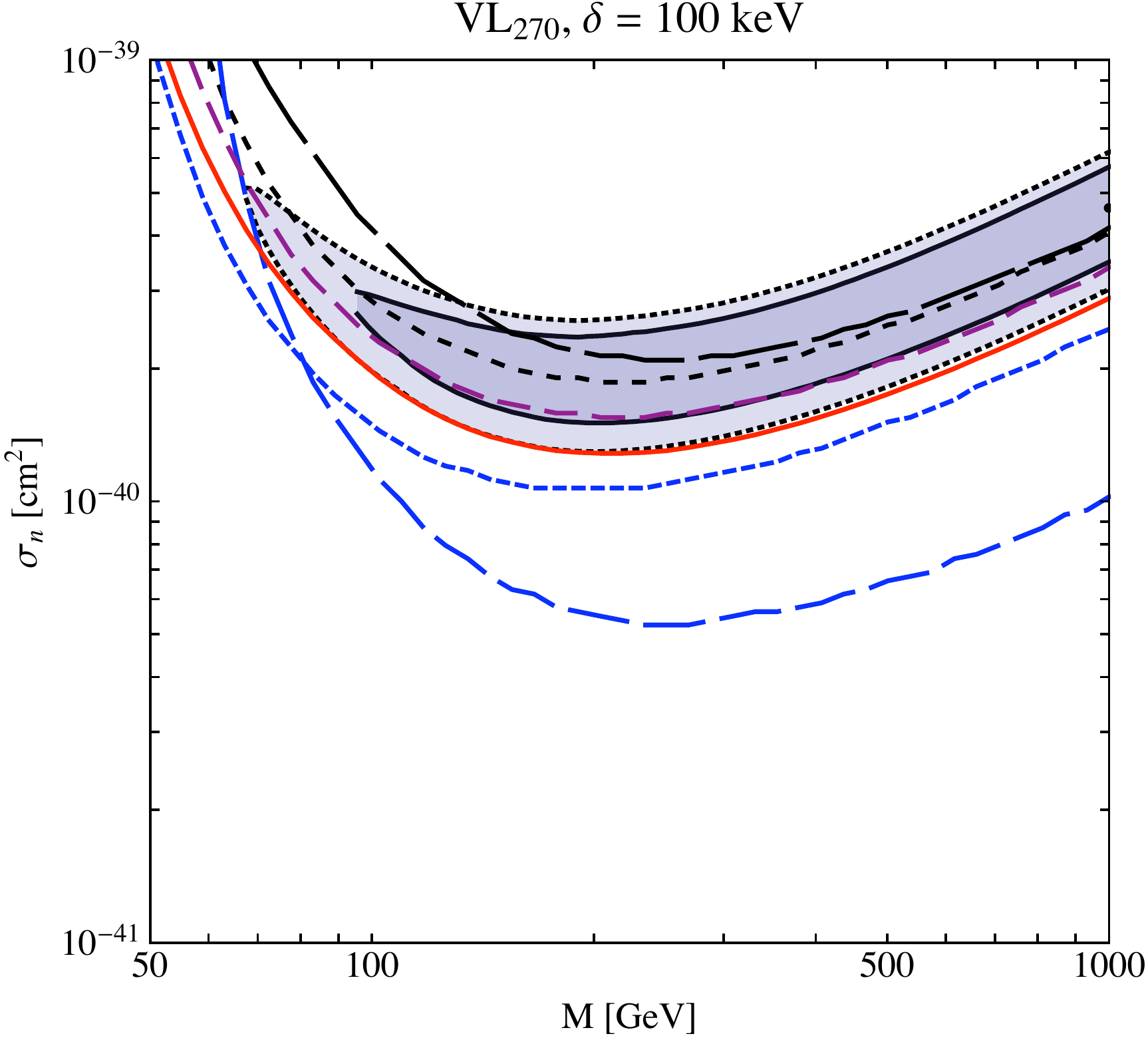}\includegraphics[height=2.8in,width=0.5\textwidth]{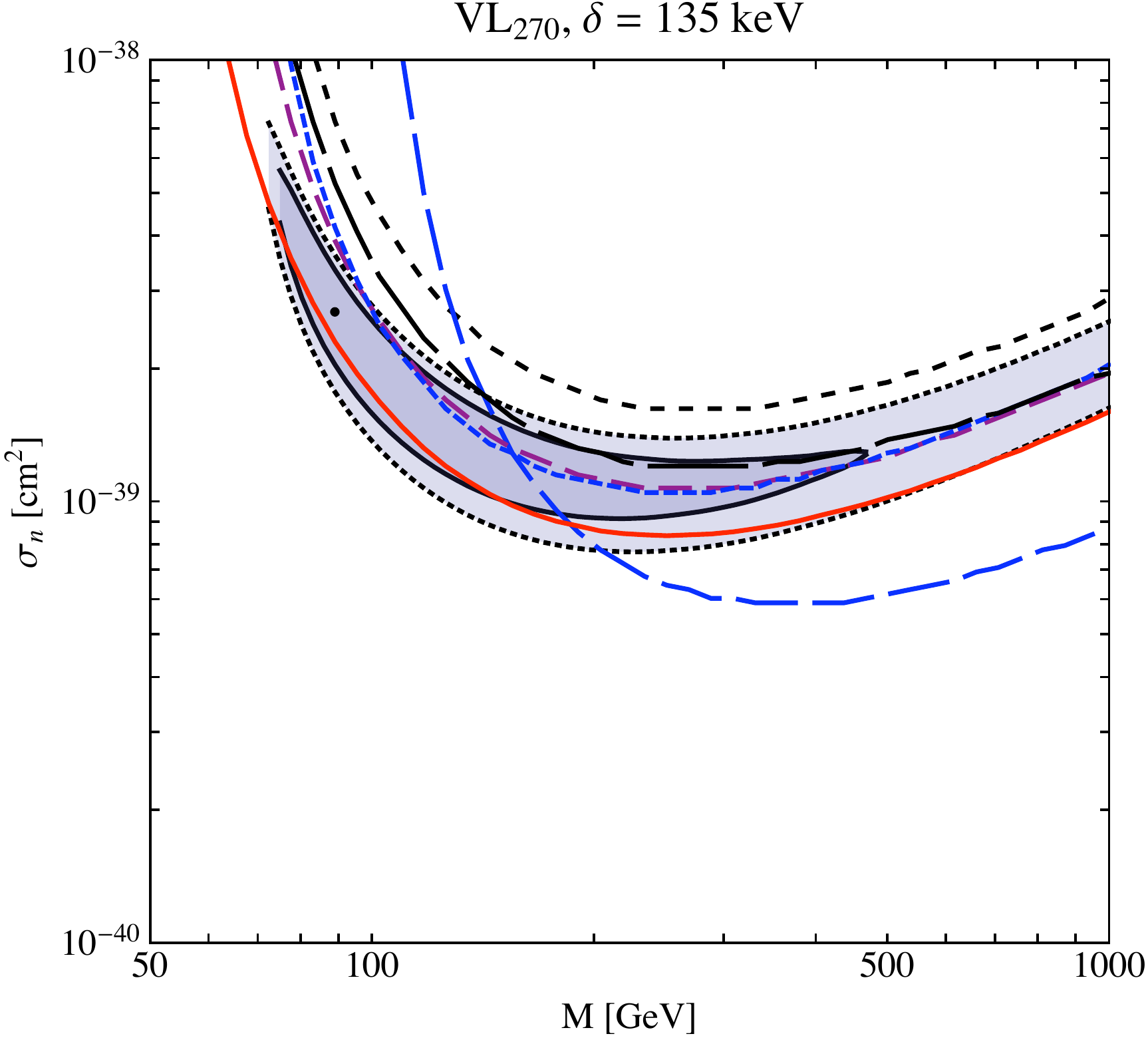}\
\caption{The allowed parameter space for fixed $\delta=100$ keV and $\delta=135$ keV while varying $M_{\chi}$. Comparing with Figure 4, we see that these results are similar to those obtained for the SHM, the difference being that the exclusion curves are typically at larger cross sections. While all of the DAMA parameter space is excluded at $\delta=100$ keV (the right panel), there is clearly some allowed parameter space at $\delta=135$ keV (the left panel) for WIMP masses less than $200$ GeV.}\label{VL270massplots}
\end{figure}

In Figures \ref{VL270deltaplots} and \ref{VL270massplots}, we present the exclusion limits for $\text{VL}_{270}$. Comparing Figure \ref{VL270deltaplots} with Figures \ref{deltaplots} and \ref{deltaplots2}, and Figure \ref{VL270massplots} with Figure \ref{massplots}, we see that although the exclusion limits from $\text{VL}_{270}$ typically occur at a larger cross section than the SHM, the DAMA allowed region is much closer to that from the SHM than $\text{VL}_{220}$.

With reference to Figure \ref{CDMSVEL}, we see that the departure from the SHM velocity distribution is greater for the $\text{VL}_{220}$ distribution compared to the $\text{VL}_{270}$ distribution, particularly in the high velocity region. However, the tail of the distribution is still much less populated in $\text{VL}_{220}$ and $\text{VL}_{270}$ compared to the SHM, which explains why the exclusion limits for $\text{VL}_{220}$ and $\text{VL}_{270}$ typically occur at larger cross sections than for the SHM.  The increased population in the $\text{VL}_{270}$ tail, relative to the $\text{VL}_{220}$ tail, and the small change in particle numbers over the region of $v_{min}$ of interest leads to stronger exclusion limits at high WIMP masses compared to $\text{VL}_{220}$.  This shows that even a small change in the dispersion can lead to relatively large changes in allowed parameter space.

\subsection{The Dark Disc and iDM}
\begin{figure}[h]
\centering
\includegraphics[height=2.8in,width=0.5\textwidth]{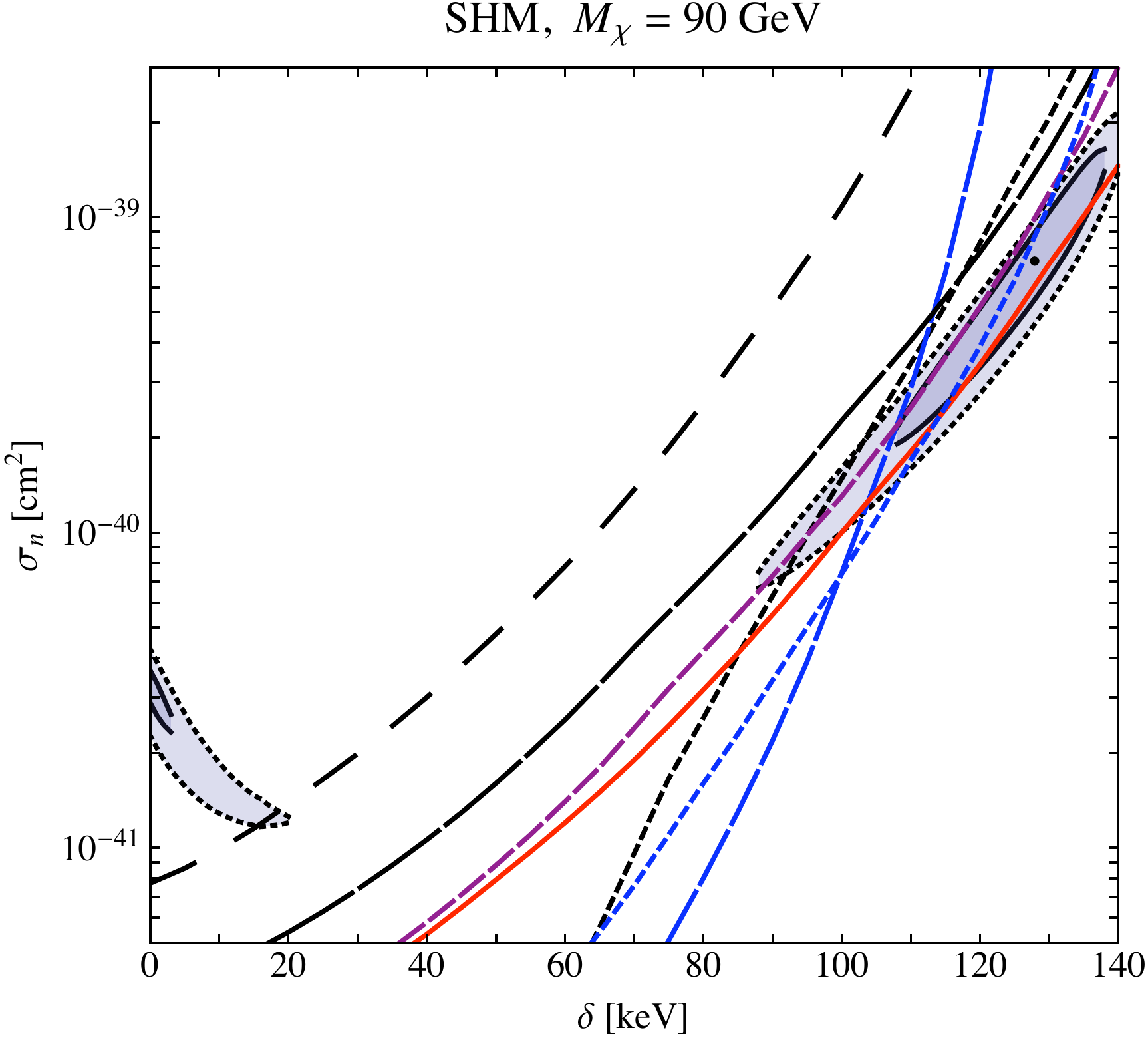}\includegraphics[height=2.8in,width=0.5\textwidth]{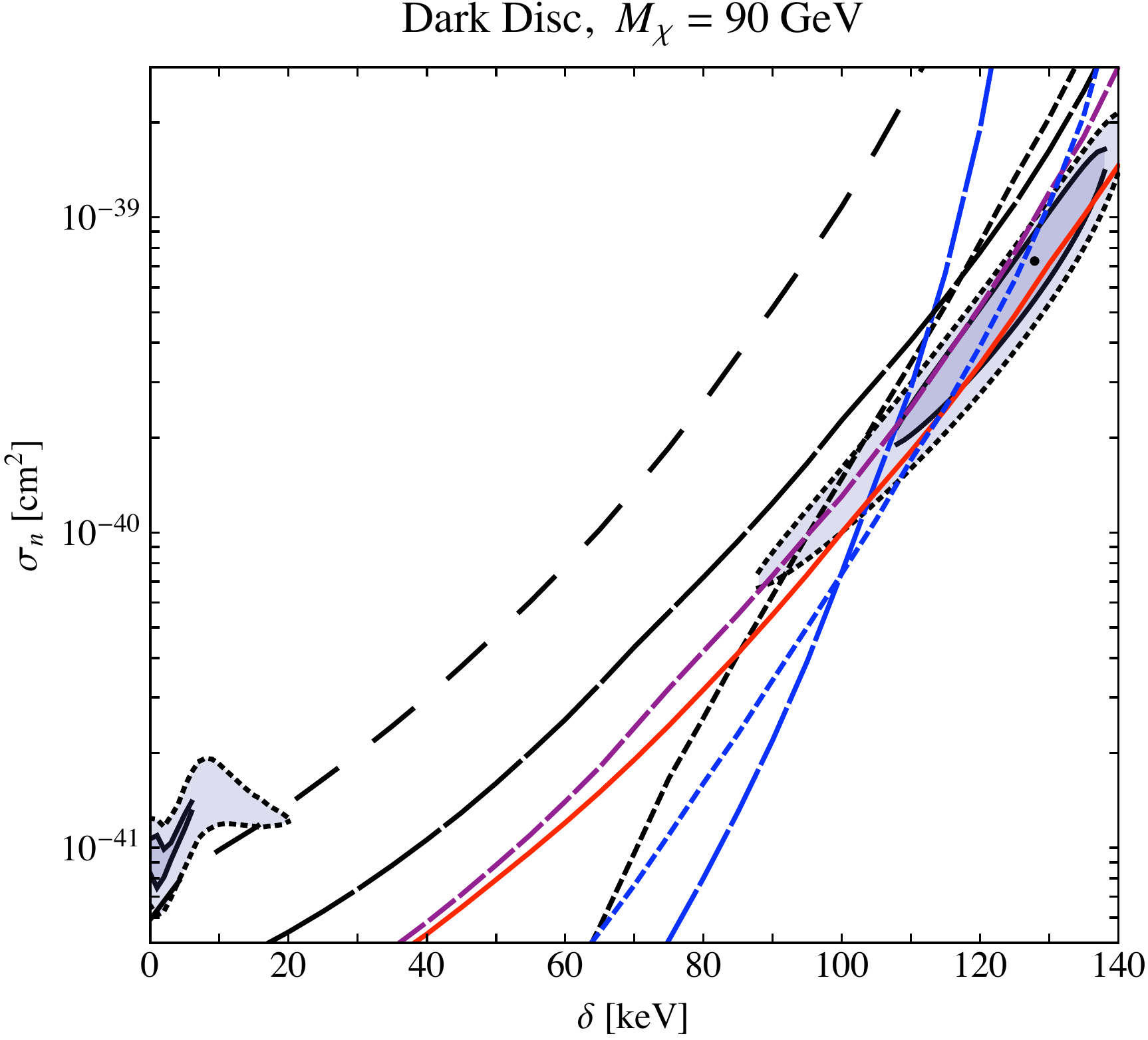}
\caption{Exclusion limits for the SHM (left panel) and the SHM + Dark Disc (right panel) for a WIMP mass of $90 \text{ GeV}$. The only major differences occur at relatively low $\delta$ where one can see the region corresponding to DAMA channeled events is deformed when the Dark Disc is included.  However, this region of parameter space is completely excluded by the other experiments.}\label{darkdisc}
\end{figure}

In Figure \ref{darkdisc} we plot the exclusion limits as a function of $\delta$ for the SHM and SHM + Dark Disc.  As a fiducial choice we have taken $R = \frac{\rho_{Disc}}{\rho_{SHM}} = 1$.
\begin{figure}[h]
\centering
\includegraphics[height=2.2in,width=0.5\textwidth]{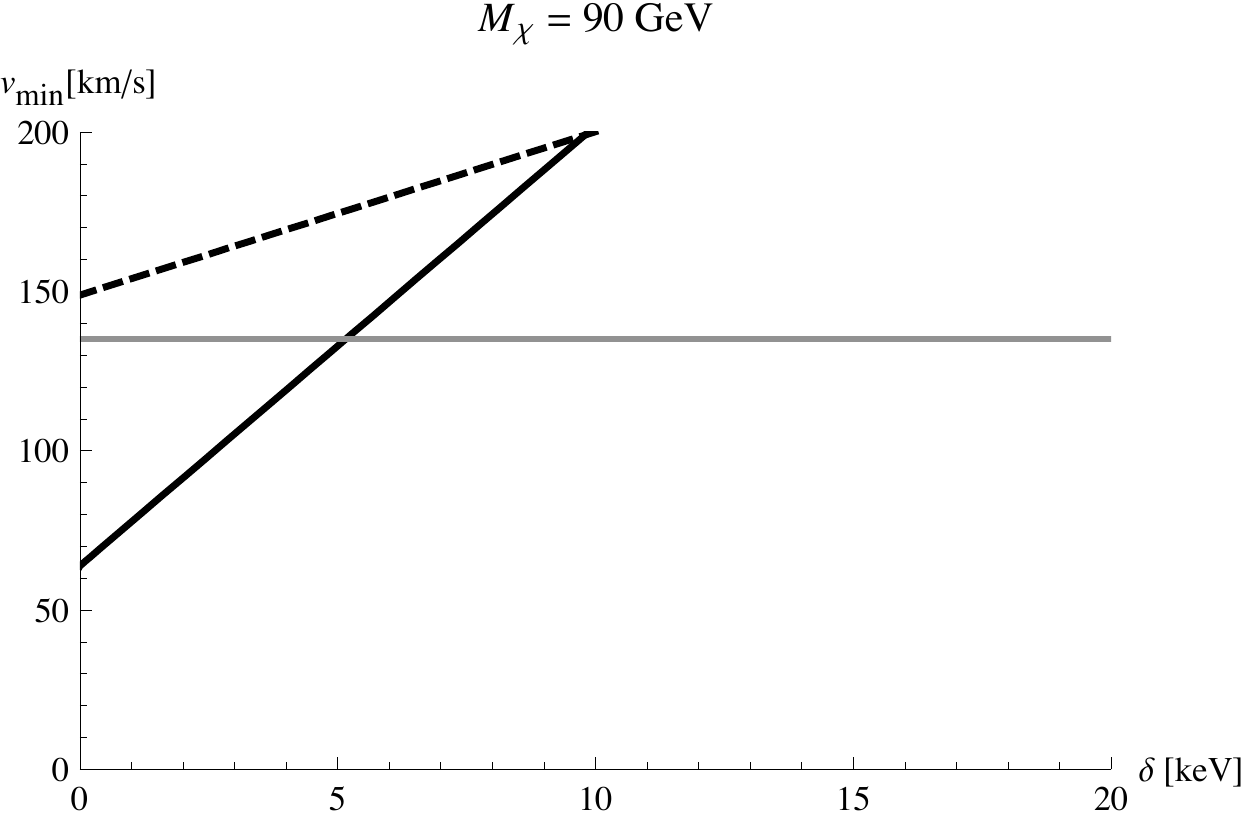}\includegraphics[height=2.2in,width=0.5\textwidth]{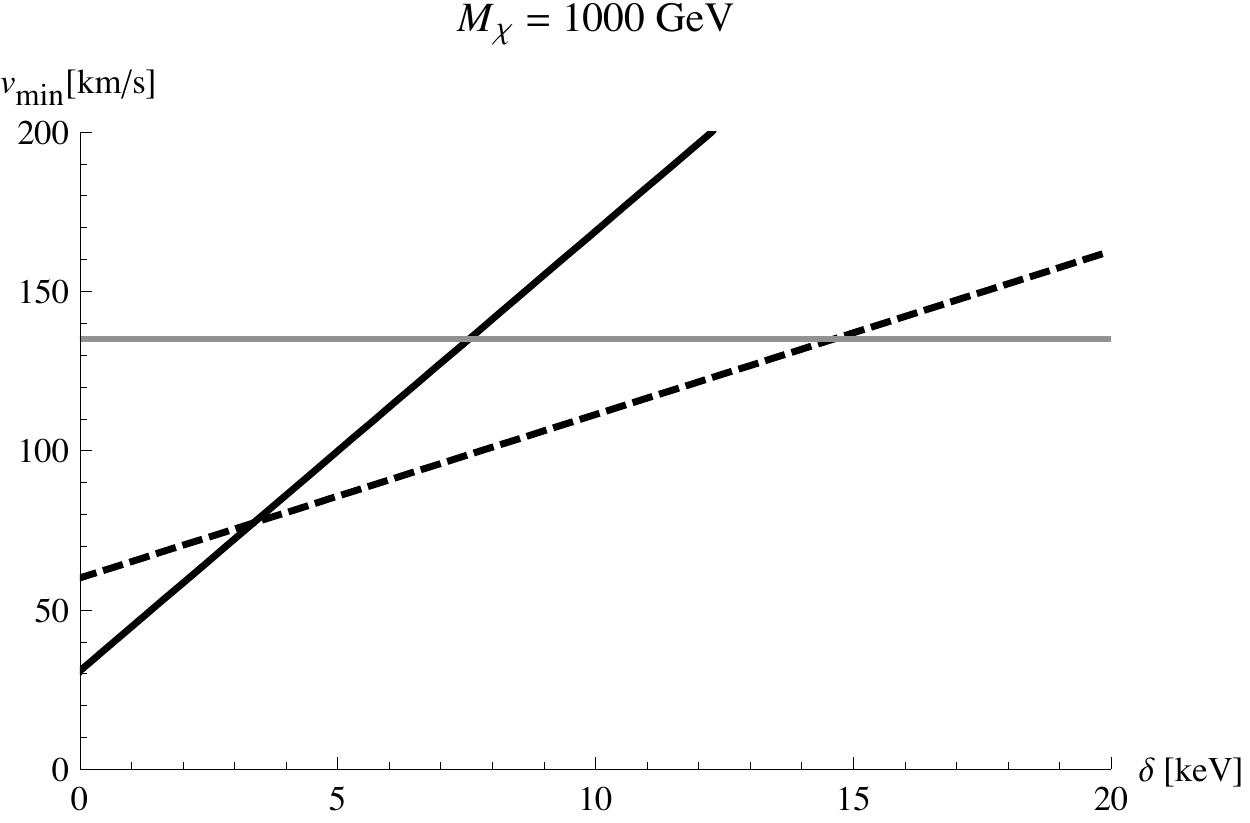}
\caption{The minimum velocity for scattering at CRESST-II (Dashed) and DAMA (Solid).  The velocities are calculated using Eq.(2.1) and taking the experimental energy threshold.  The DAMA line corresponds to channeled scattering and the gray line is roughly the maximum velocity at which the Dark Disk particles would be found. For $\delta\ge 15$ keV very few particles in the dark disc will scatter.}\label{minv}
\end{figure}

One can immediately see that the dark disc only influences the limits at very low $\delta$.  Increasing the density of the dark disc will change the limits at low $\delta$ but not the value of $\delta$ at which the disc is detectable.  The fact that the dark disc is not detectable at high $\delta$ is due to the increased minimum velocity required for scattering.  As the disc lags the motion of the Sun by 40 km/s, and has velocity dispersion of $39$ km/s in the tangential direction, 99\% of disc particles will only have a relative velocity up to $v_{DD} \sim 135$ km/s.  If the minimum velocity for scattering is above this the dark disc will have little effect on event rates.  This effect is demonstrated in Figure \ref{minv}.

One can see that for heavier WIMP masses the effect of the Dark Disc would increase, but would still only be found for $\delta < 20$ keV.  These results would suggest that possible sub-halo components which are rotating roughly in the same way as the visible galactic plane would have little or no effect on inelastic scattering event rates, whereas the effect on elastic scattering rates should be pronounced.  This effect could have interesting consequences for clumpy dark matter.

\subsection{Iodine quenching factor} \label{quench}
\begin{figure}[h!]
\centering
\includegraphics[height=2.8in,width=0.5\textwidth]{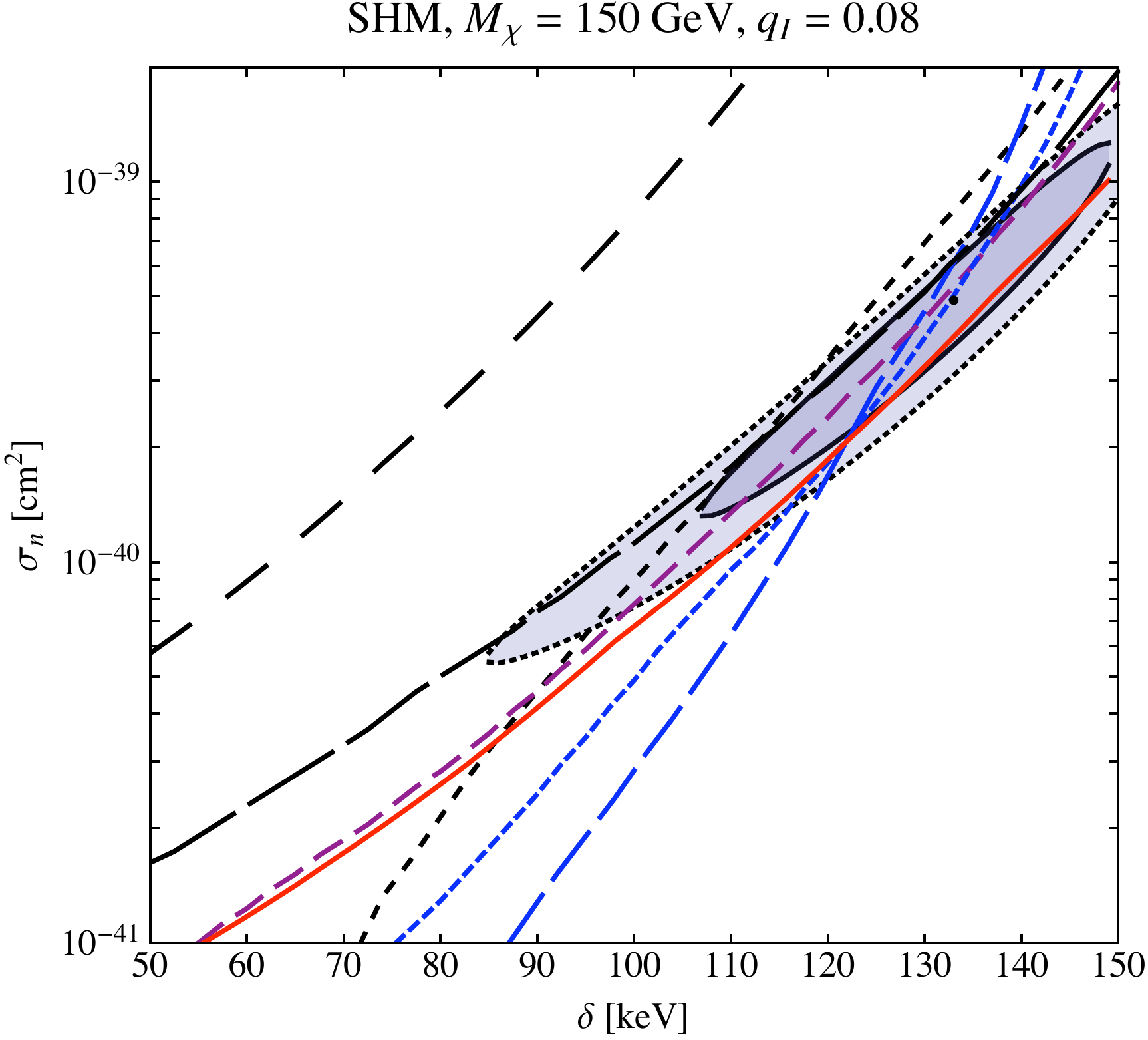}\includegraphics[height=2.8in,width=0.5\textwidth]{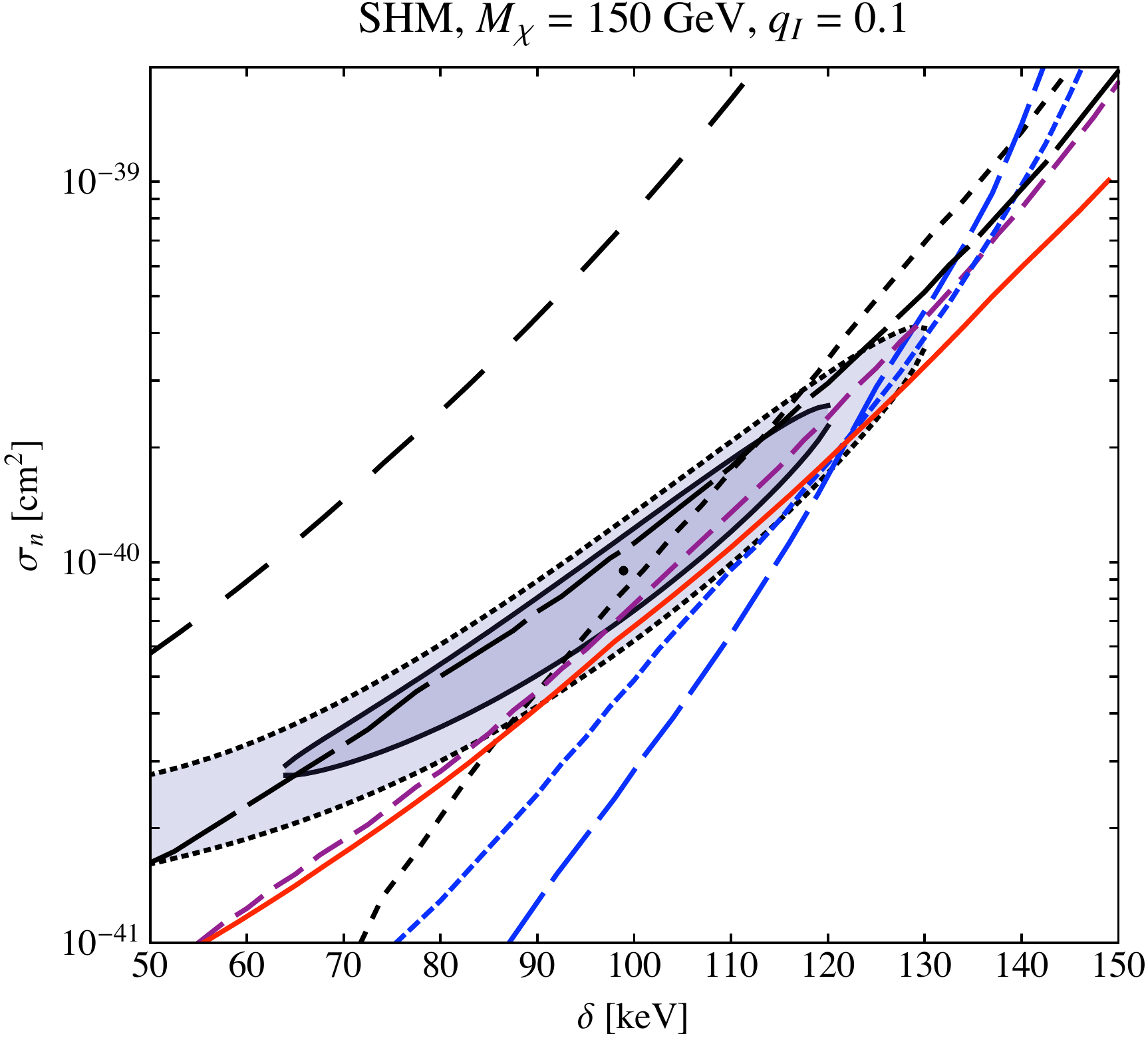}
\caption{The exclusion limits with an iodine quenching factor of $q_I = 0.08$ (left panel) and $q_I = 0.1$ (right panel), for a WIMP mass of $150 \text{ GeV}$ and using the SHM. One can see that the preferred region of parameter space for the DAMA results can move by $\Delta(\delta) \sim 30$ keV and $\Delta(\sigma_n) \sim 5 \times 10^{-40} \text{cm}^2$, leading to agreement with or exclusion by the other experiments.}\label{quenchfig}
\end{figure}

In \cite{DAMA:quench} the iodine quenching factor is given as $q_I = 0.09 \pm 0.01$.  As  changing the value of the quenching factor will stretch the recoil spectrum as a function of recoil energy it is interesting to see what effect varying the quenching factor has on the DAMA/LIBRA preferred region of parameter space, and whether this strongly influences the agreement with other experiments.  The effect of this variation is illustrated in Figure \ref{quenchfig}.

As one can see, the overall effect can be significant, and the DAMA/LIBRA preferred region can shift by a relatively large amount.  For the lower quenching factor there is a region of agreement between all experiments at the 90\% confidence level, whereas for the higher quenching factor there is disagreement between DAMA/LIBRA and the CRESST-II and CDMS II experiments at almost a 99.5\% level for the SHM.

As this effect is so pronounced it would be interesting  if the DAMA collaboration are able to put tighter bounds on the quenching factors in their experiment.  For elastic scattering the uncertainty is not so significant as it is the channeled event regions where $q_I = 1$ that show best agreement with the other experiments.  However, as inelastic scattering finds greatest agreement for the region of parameter space corresponding to quenched events, the value of the quenching factor is of much greater importance.

\subsection{Circular velocity} \label{circular}
\begin{figure}[h]
\centering
\includegraphics[height=2.8in,width=0.5\textwidth]{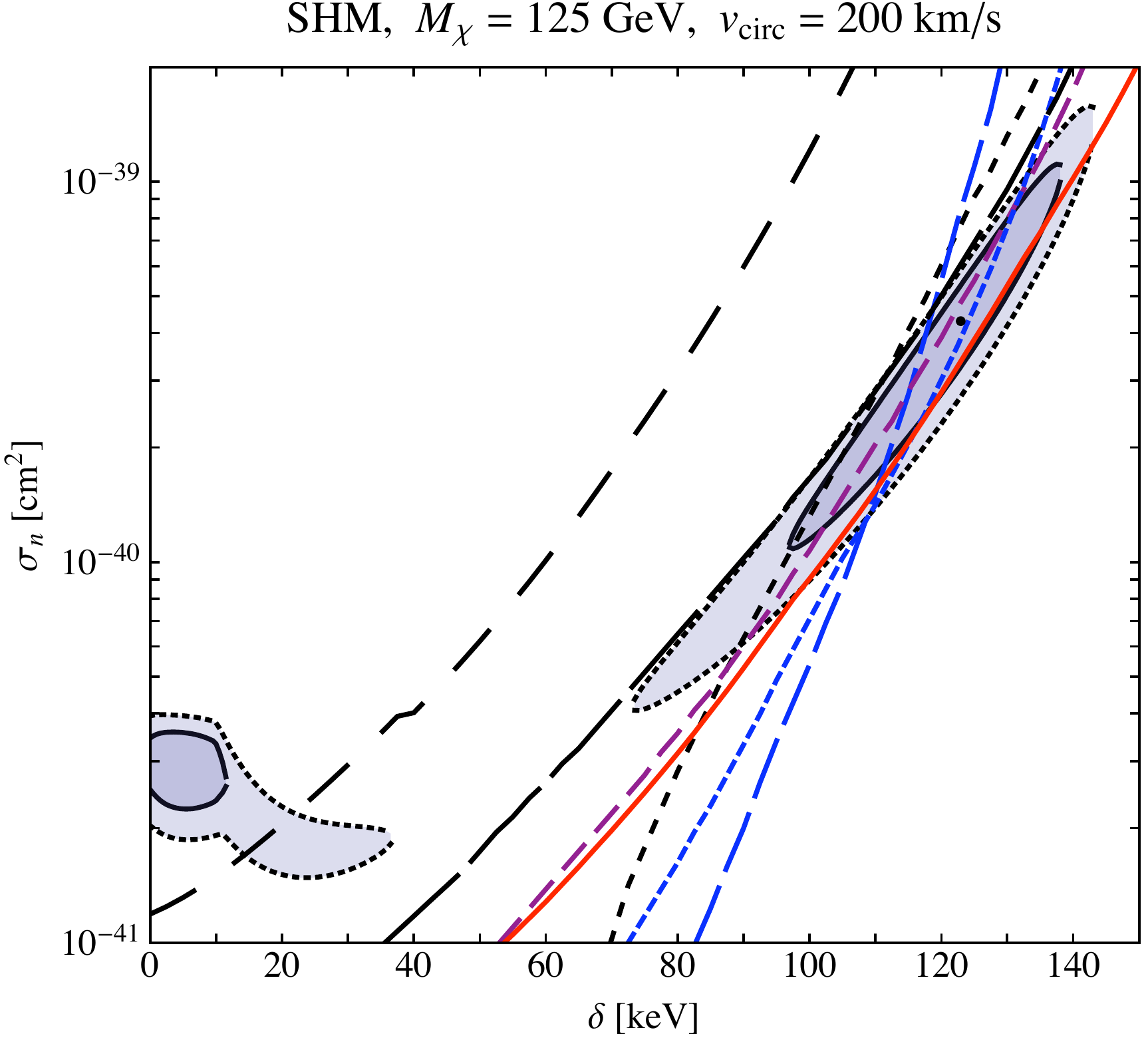}\includegraphics[height=2.8in,width=0.5\textwidth]{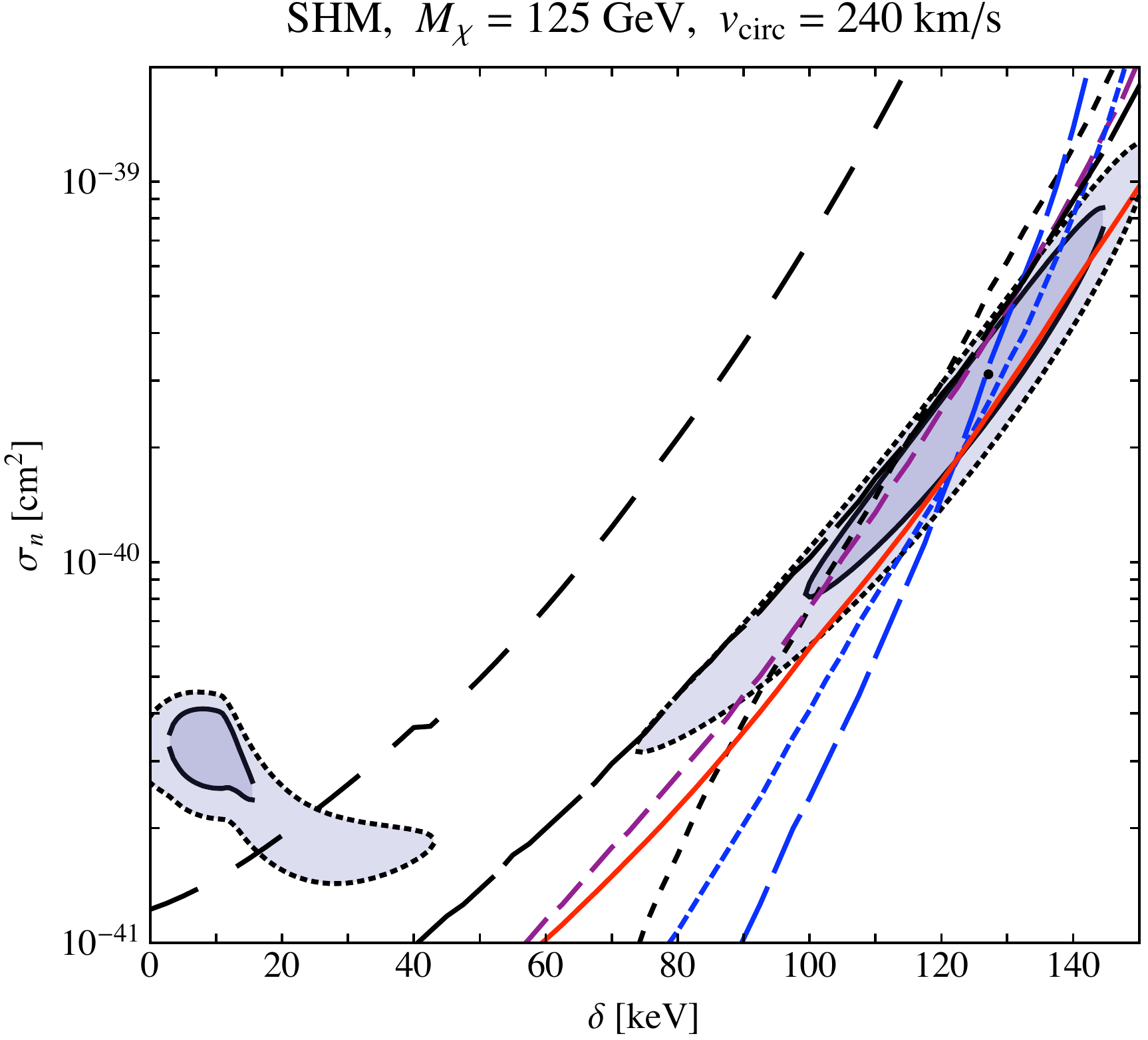}
\caption{The exclusion limits for a a local circular velocity of $v_{circ} = 200$ (left panel) and $v_{circ} = 220$ (right panel), for a WIMP mass of $125 \text{ GeV}$ and using the SHM. One can see that the limits change a small amount and that the overall agreement with the DAMA results is slightly better for a lower circular velocity.  However, the overall qualitative effect of changing the circular velocity is relatively small.}\label{circ}
\end{figure}

In  \cite{BinneyBook} the Sun's circular velocity about the center of the Milky Way is given as $v_{circ} = 220 \pm 20 \text{km/s}$.  In the iDM scenario the majority of observed events are expected to be from scattering off WIMPs at the high end of the tangential velocity distribution, so it is expected that varying the circular velocity will have an effect on the exclusion limits set by detectors.  In Figure \ref{circ} we plot the exclusion curves for a WIMP of mass $M_{\chi} = 125$ GeV for $v_{circ} = 200, 240 \text{km/s}$.

As expected this variation does affect the exclusion limits, however there is little overall change in the qualitative features.  Unlike with the Dark Disc this variation occurs over all ranges of $\delta$.  Due to the enhancement of the signal, through increasing the number of particles with $v > v_{min}$, increasing the circular velocity of the Sun generally leads to marginally stronger upper limits on cross sections for iDM.

\subsection{Local galactic escape velocity}\label{escapevel}
\begin{figure}[h]
\centering
\includegraphics[height=2.8in,width=0.5\textwidth]{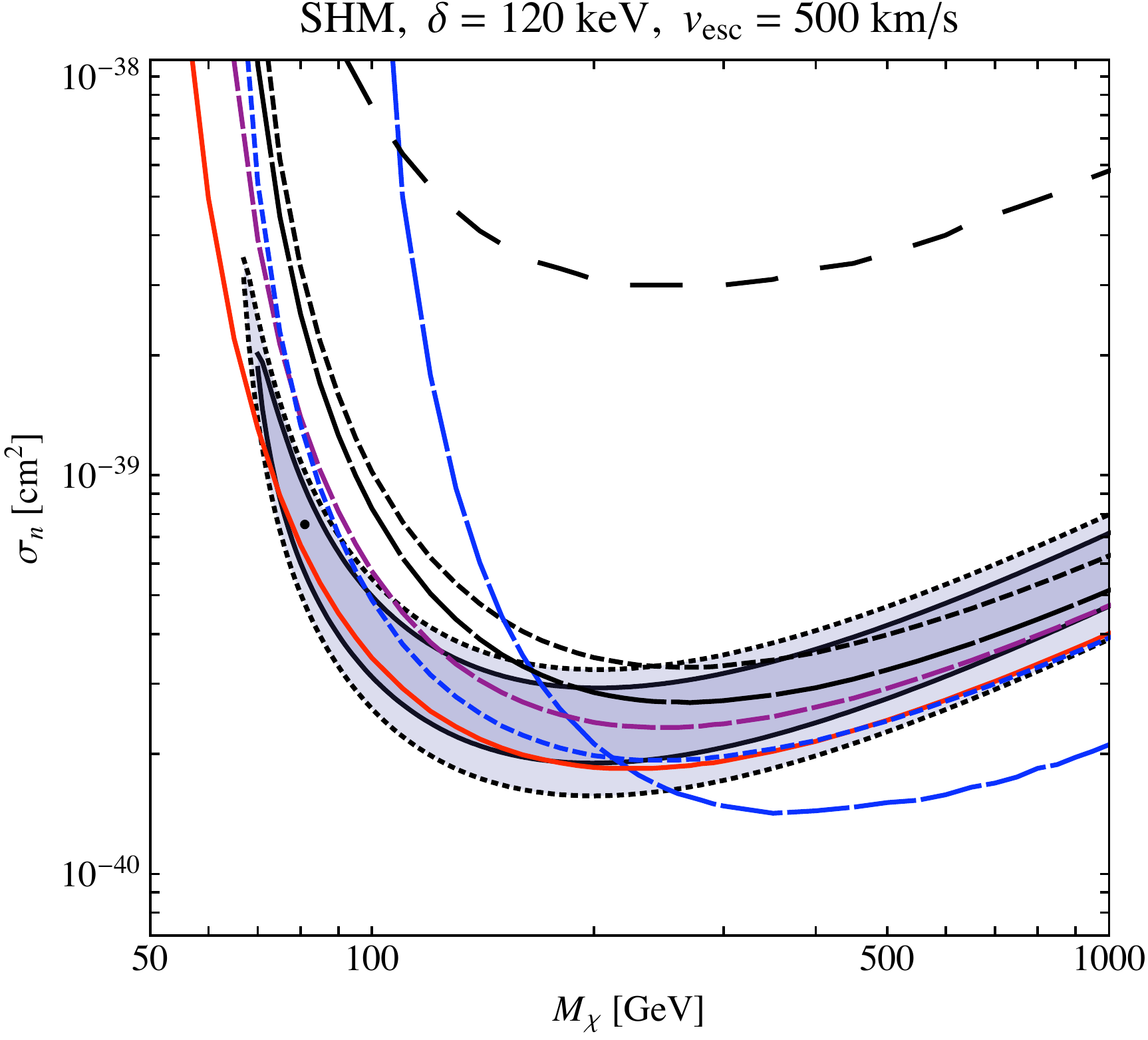}\includegraphics[height=2.8in,width=0.5\textwidth]{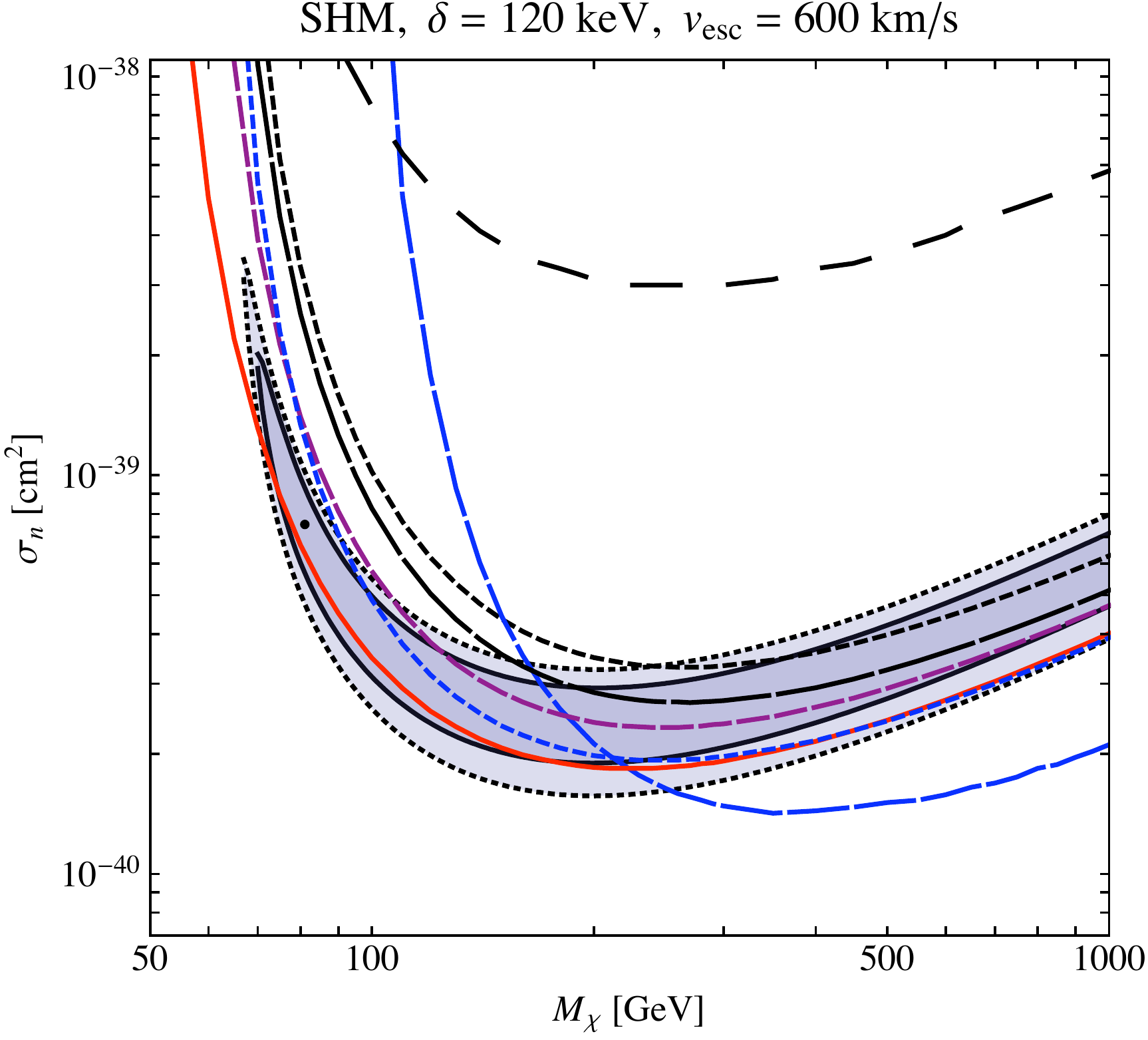}
\caption{Exclusion limits for the SHM for $v_{esc} = 500$ km/s (left panel) and $v_{esc} = 600$ km/s (right panel).  There is better agreement between experiments for lower escape velocity and the limits from CDMS II show the most sensitivity to the escape velocity.}\label{escape}
\end{figure}

The RAVE survey have measured the escape velocity to lie within the range $498 < v_{esc} < 608 $ km/s at $90\%$ confidence \cite{Smith:2006ym}. In this paper, we have taken the fiducial value $v_{esc} = 550$ km/s, however the experiments are only sensitive to the tails of the velocity distribution, which depend on the escape velocity. In Figure \ref{escape} we illustrate what the effect of varying the escape velocity within the range allowed by the RAVE survey has on the experimental limits.

The limit which shows the largest change is from CDMS II, although all experimental limits show some variation, particularly at lower masses. This effect can be understood with reference to Figure \ref{CDMSVEL}. At lower masses, the minimum velocity is closer to the escape velocity, therefore for lower escape velocities, the range of integration is smaller, hence the limits are weaker. At higher masses, the relative change in varying the escape velocity is lower because the range of integration is larger, so the difference for all experiments is smaller.

\section{Conclusions} \label{Conclusions}

We have shown, in the context of iDM, that the region of agreement between the DAMA data and results from other experiments is sensitive to the uncertainties present in the galactic WIMP velocity distribution.  In particular we have found that the other direct detection experiments,
including the most recent ZEPLIN-II and III, CRESST-II, XENON10, KIMS, and CDMS-II data sets, do not exclude the region of parameter space preferred by the DAMA results up to WIMP masses $M_\chi \sim \text{1 TeV}$ when the $\text{VL}_{220}$ velocity distribution is used (see Figures \ref{deltaplots2} and \ref{massplots}), while for the $\text{VL}_{270}$ velocity distribution, WIMP masses $M_\chi \leq \text{200 GeV}$ are allowed (see Figures \ref{VL270deltaplots} and \ref{VL270massplots}).  Furthermore, we have also argued that the region of agreement between experiments is very sensitive to the quenching factor used to interpret the DAMA data (see Figure \ref{quenchfig}), and also to the local galactic escape velocity (see Figure \ref{escape}), both of which presently have $\sim$10\% uncertainties, and somewhat to the Sun's circular velocity (see Figure \ref{circ}).

Independent of experimental set-ups we have also shown that, in the iDM scenario, detectors would be insensitive to WIMPs traveling with a small velocity relative to the sun due to the increased minimum velocity for scattering.  This would lead to dark matter clumps or streams being undetectable if rotating in the galactic plane, as in the Dark Disc halo model, or streaming with a small relative velocity (see Figure \ref{darkdisc}).

New data from heavy element experiments such as CRESST-II and XENON100 could have the potential to completely exclude the DAMA preferred region for iDM, especially if taken during the modulation maximum.  In particular, if the planned EURECA experiment were to use a tungsten detector it would provide significant limits on both the elastic and iDM scattering cross-sections.  Furthermore because of the low energy cut-off in the recoil energy spectrum for iDM, it is important that the experiments increase their sensitivity at higher recoil energies, where the iDM recoil spectrum peaks.

Without a better understanding of the details of the velocity distribution in our galaxy, including the effects of the baryons, precise statements about the consistency of various direct detection experiments are not possible.
\\

\noindent{{\bf Note added:} Subsequent to the submission of this work two recent DM simulations were brought to our attention:

\begin{itemize}
\item  Recently the results of a second simulation, Via Lactea II \cite{Kuhlen:2008qj}, were made public.  These results showed further detail in the substructure of the DM halo, with the clumpiness extending over six orders of magnitude, down to the resolution limit.  

\item  Another Milky Way-sized DM halo simulation, the Aquarius Project, also recently released their results \cite{Springel:2008cc}, and the phase-space structure of these simulated halos was the subject of further study \cite{Vogelsberger:2008qb}.  The velocity distributions found varied from simulation to simulation and showed interesting bumpy features.  Further to this, in \cite{Vogelsberger:2008qb} the influence of these velocity distributions on direct detection signals for elastically scattering dark matter was investigated and it was found that the recoil spectra could deviate by up to 10\% from that expected from the best-fit multivariate Gaussian.  

\end{itemize}

These simulations provide further strong support for deviations from the SHM, but as data from these simulations are not (to the authors knowledge) publicly available, we have not been able to update our study to include them, although we hope to return to this subject in a future publication.

\section{Acknowledgements}
We would like to thank James Binney, Malcolm Fairbairn, Graham Ross, Joe Silk, and Nigel Smith for discussions, Neal Weiner for assistance comparing his earlier analyses with ours, Lawrence Hall for discussions and for informing us of Ref.\cite{Fairbairn:2008gz}, and especially, Hans Kraus, Hyun Su Lee, and Kaixuan Ni, for providing us with experimental data and for discussions.   We also wish to thank Michael Kuhlen for bringing subtleties regarding the Via Lactea gravitational potential to our attention.  CM and MM are supported by STFC Postgraduate Studentships. This work is partially supported by the EC network 6th Framework Programme Research and Training Network Quest for Unification (MRTN-CT-2004-503369) and by the EU FP6 Marie Curie Research and Training Network UniverseNet (MPRN-CT-2006-035863).

\bibliographystyle{JHEP}
\bibliography{myrefs}

\end{document}